\documentclass[sigconf,nonacm]{acmart}

\usepackage[utf8]{inputenc}
\usepackage{csquotes}
\usepackage{enumitem}
\usepackage[hyphenbreaks]{breakurl}

\title[Mutual Benefit]{Mutual Benefit: The Case for Sharing Autonomous Vehicle Data with the Public}

\begin{document}

\author{David Goedicke}
\affiliation{%
  \institution{University of Duisburg-Essen}
  \city{Essen}
  \country{DE}
}
\affiliation{%
  \institution{Cornell Tech}
  \city{New York}
  \country{US}
}

\author{Natalie Chyi}
\affiliation{%
  \institution{Cornell Tech}
  \city{New York}
  \country{US}
}

\author{Alexandra Bremers}
\affiliation{%
  \institution{Cornell Tech}
  \city{New York}
  \country{US}
}
\author{Stacey Li}
\affiliation{%
  \institution{Cornell Tech}
  \city{New York}
  \country{US}
}
\author{James Grimmelmann}
\affiliation{%
  \institution{Cornell Tech}
  \city{New York}
  \country{US}
}
\author{Wendy Ju}
\affiliation{%
  \institution{Cornell Tech}
  \city{New York}
  \country{US}
}

\begin{abstract}
Autonomous driving is a widely researched technology that is frequently tested on public roads. The data generated from these tests represent an essential competitive element for the respective companies moving this technology forward. In this paper, we argue for the normative idea that a part of this data should more explicitly benefit the general public by sharing it through a trusted entity as a form of compensation and control for the communities that are being experimented upon. To support this argument, we highlight what data is available to be shared, make the ethical case for sharing autonomous vehicle data, present case studies in how AV data is currently shared, draw from existing data-sharing platforms from similar transportation industries 
to make recommendations on how data should be shared and conclude with arguments as to why such data-sharing should be encouraged. 
\end{abstract}

\maketitle

\section{Introduction}
The advent of autonomous vehicles, and the development and testing of those vehicles in public spaces, raise key ethical issues about ownership and sharing of data. In this paper, we make a philosophical argument for sharing data collected about the public by privately owned autonomous vehicle companies. This is an important issue for the CHI community, because of the increasing interest in vehicle automation within the community \cite{kun2016shifting, ayoub2019from}, and our interest in the data and subsequent research it would enable \cite{mirnig2019insurers, riener2014collective}. Moreover, it builds upon prior arguments in the CHI community around community informatics\cite{FiveProvocations}, which posits a moral case that the people whose activity and experiences will ultimately be affected most directly by a design outcome ought to have a substantive say in what that outcome is \cite{carroll2007participatory}.

We argue that the use of public roads and the risk to public persons for the development of intellectual property for private entities demands some immediate benefit in exchange. In this paper, we build the case for this position. First, we describe these on-road autonomous vehicle experiments, discuss what data is being collected and why, make an argument that the data from these deployments should be shared, and discuss a mechanism by which data could be shared. This data sharing would help to reduce the risk associated with multiple companies duplicating similar public tests and provides broader insights that can benefit the wider community. 

 This is an important issue for the CHI community, both because of the increasing interest in vehicle automation \cite{kun2016shifting, ayoub2019from}, but also our interest in the data and subsequent research it would enable \cite{mirnig2019insurers, riener2014collective}. In a nutshell, we argue that the use of public roads and the risk to public persons for the development of intellectual property for private entities demands some immediate benefit in exchange. 

To make this case, we 
enumerate costs and risks associated with these deployments, consider arguments for and against different types of data sharing, and make policy recommendations. We further weigh and respond to legal challenges around the sharing and recommendations for the US and EU jurisdictions.  

This paper contributes an argument for AV data sharing, which empowers the CHI community to advocate sharing data in any jurisdiction considering autonomous vehicle deployments. Furthermore, we hope that the case of data sharing in autonomous vehicles can inform larger ethical discussions about data sharing, which we foresee will become possible or necessary as data-driven "intelligent" products and services become increasingly common.

\subsection{Why is on-road data collection occurring?}

 While it seems inevitable that autonomous vehicles will eventually become commonplace, it has been acknowledged that the arrival of a truly autonomous car---the driverless car that can go anywhere, anytime---
  is a long way off, with some forecasters predicting that safe, widespread adoption 
 of autonomous passenger vehicles
 will take decades.\cite{gessner_2020} 
The autonomous passenger cars that people have spotted in their neighborhoods or witnessed on YouTube are actually geographically-limited experimental deployments of key technologies being developed and tested.

Much of what is occurring on-road currently is actually data collection--about road conditions, traffic interactions, pedestrian behavior, and passenger experience--that are needed to enable the algorithms and control models that future autonomous vehicles will use. While some testing and development can occur on closed tracks and in simulation, on-road testing provides key information and lessons that are an irreplaceable step toward the realization of autonomous vehicles. Simulations alone are inadequate to the range of real-life situations AVs will encounter: no plausible simulation could have incorporated a ``woman in [an] electric wheelchair chasing a duck with a broom'' scenario witnessed by one GoogleX (now Waymo) car. \cite{mercury_2016}  Still, as unusual as situations may be that AVs will encounter on the road, they will be expected to deal with all of them safely and appropriately. In order for AVs to be sophisticated enough to be widely adopted and safe enough to use, then, they will have to be tested on public streets, with some risk to those in their path.

We propose that one way to mitigate this risk and to build safer AV technology faster is through mandatory sharing of safety data by AV companies when they are testing their vehicles. In particular, we believe that they should be required to share detailed information about their operation of test vehicles with increased published data for any in disengagement, near-crash, and crash incidents. Furthermore, publicly sharing data from these deployments can benefit research on user interactions with transportation and municipalities in ways that are beneficial to human-computer interaction researchers.

\section{Related Work}
Within the HCI community, we have seen some focus on \textit{legal} issues \cite{mirnig2019insurers, inners2017beyond}, but seldom considered \textit{policy}--not what is \textit{allowed}, but what should be. Partly this is due to the fact that the policymaking activity takes place within bodies such as the Society of Automotive Engineers (SAE), the International Organization for Standardization (ISO), and policy study committees of organizations such as the Transportation Review Board (TRB), and occur outside of the public, academic discussion. These organizations have slightly differing agendas. Engineering bodies develop standards to enable commonality in nomenclature,  discourse, and system characteristics. Standards bodies often perform certification to ensure the quality, safety, and reliability of products, systems, and services. Policymakers for governments and municipalities often have a more complicated task of balancing the competing interests of industry and the public. Nevertheless, we believe the CHI community should consider and even advocate policy since issues such as data sharing and public testing of AVs have important implications for HCI research \cite{braun2018automotive}. 
This type of data sharing has been important to the flight automation community both for safety and research \cite{halford2008asias}.

To date, in the US, the National Highway Traffic Safety Administration (NHTSA) has explicitly avoided making direct policy and has been relatively circumspect in its advice to states for their own AV policies. This is probably because the technology in question is still nascent, and NHSTA is waiting to see how it develops. The 2013 ``Preliminary Statement of Policy Concerning Automated Vehicles'' \cite{nhtsa2013preliminary} primarily recommended that states permit the operation of self-driving vehicles only for testing. NHSTA's guidance to date  \cite{nhtsa2021website} has been to license the safety drivers to make sure they are able to operate the AV safely, that the state on-road testing to minimize risk to other road users, that there be a safety driver capable of take over if required for any vehicle being tested, that they limit testing to environmental conditions suitable to the AV, and that they establish reporting requirements to establish the performance of the AV during testing. 

There is scant work exploring what kinds of data should be reported except for crash and critical incident data. Some, such as Claybrook and Kildare, \cite{claybrook2018autonomous}, argue that mandatory data should be reported but focus on requiring detailed data on crashes and system malfunctions. 
``Currently, there is no transparency regarding the algorithms that form the basis for AV function and thus no way to determine whether there are better approaches to solving problems that resulted in collisions or serious system malfunctions. The information required, however, is more than just that covered by an incident report but must include details on the dynamics of the collision and, more importantly, how the decision process of the AV may have led to or contributed to the crash. Only through data collection and analysis can future regulatory needs be developed and justified. '' They propose as a model the Aviation Safety Information Analysis and Sharing (ASIAS) Program \cite{asias_factsheet}. (More detailed discussion of the ASIAS program and system are described later in this paper.) 

From an HCI perspective, we believe that detailed data about the road interactions of AVs  and the context in which those interactions occurred is also relevant and critical to the public, as it informs the qualitative manner in which the vehicle interacts with the public. Ethnographic research from the human factors community have noted the ways in which social interaction occurs between drivers \cite{juhlin1999traffic}, driver-bus interaction \cite{normark2006enacting}, pedestrian-vehicle interaction \cite{vsucha2014road, dey2017pedestrian} and interactions at petrol stations \cite{normark2006tending}, and AV interactions with other road users \cite{vinkhuyzen2016developing, brown2017social}. 
More recent research has made use of large-scale naturalistic driving datasets such as Virginia Tech’s 100 car Naturalistic Driving Study  \cite{dingus2006100}, the follow-on SHRP2 Naturalistic Driving Study \cite{hankey2016description}, University of Minnesota’s Field Operational Test of the Teen Driver Support System \cite{creaser2015teen}, and MIT’s Autonomous Vehicle Technology Study \cite{fridman2017autonomous} to better understand how vehicles interact with each other, as well as with pedestrians and other road users, \cite{fridman2019advanced, domeyer2020interdependence}. This type of data critically informs our future with computer-controlled vehicles, but, to date, has been considered the private information belonging solely to the automotive companies running test AV deployments.

This paper is the first to make a case for data-sharing policies around the ordinary operation of AVs. Due to length constraints, this paper is primarily written towards influencing the laws, policies, and regulations in the United States; however, this argument is informed by policy and deployment examples from other countries, based on the limited data publicly available, and considers, in particular, the application of the EU's General Data Protection Regulation (GDPR) to the data sharing we advocate; we imagine this work might be the basis for a larger body of work considering how AV data sharing might work worldwide.


 
\section{Why should AV data be shared?}
\label{sec:why}

\textit{Why} should data from autonomous vehicle deployments be shared? Research ethics suggests that some form of data sharing may be necessary to overcome the ethical problems of subjecting the public to the risks of AV testing.  Data sharing is both a proxy (albeit an imperfect one) for true informed consent and a way of ensuring that the benefits of AV testing are as broadly distributed as the risks.

    Recently, an engineer at Uber's self-driving-car unit likened their vehicles to "a science experiment" \cite{bort2019experiment}, which is an apt description. As it stands, companies testing their AVs on public roads are essentially conducting an experiment on anyone they cross paths with or without their knowledge or consent.  AV testing meets the definition of `research' in the Federal Policy for the Protection of Human Subjects, better known as the Common Rule, the principal regulation applicable to federally-funded research: it is a  ``systematic investigation ... designed to develop or contribute to generalizable knowledge'' \cite{commonrule} \footnote{Because most AV testing is not federally funded, the Common Rule does not directly apply to it.  We draw on it as a source of widely accepted ethical norms, not as a source of law.}  Members of the public who interact with AVs, including pedestrians and other drivers, may or may not be ``human subjects'' in the sense the Common Rule uses the term: they are ``living individual[s] about whom an investigator ... [o]btains information ... through intervention,''  where ``intervention'' is defined to include ``manipulations of the ... subject's environment.'' \cite{commonrule} But even if these members of the public are not themselves considered research participants, AV testing is a research program that exposes them to risks and as such raises the same ethical issues.

Fortunately, there is a widely used framework for working through these ethical issues.  While AV testing is different in some ways from the biomedical research (e.g. vaccines and medical devices) at the heart of human-participant research regulation, the same basic moral framework and ethical principles are applicable to assessing any research with the potential to harm participants.  It was first outlined in the Belmont Report\cite{belmont1979} and later codified in the Common Rule in 1991. Twenty U.S. agencies and departments currently require Common Rule protections for research participants in research they conduct or fund, including the Consumer Product Safety Commission and Department of Transportation.

The Belmont Report presents three basic principles: Respect for Persons, Beneficence, and Justice.  Each has implications for AV testing.

\textbf{Respect for Persons}, which implements a deontological principle of treating people as ends rather than merely as means, usually requires that `` individuals should be treated as autonomous agents'' and thus that they ``enter into the research voluntarily and with adequate information,'' i.e. by giving genuinely informed consent to participate, a requirement spelled out in more detail in the Common Rule at 45 CFR §~46.116. This principle is more straightforward to uphold in clinical trials, online surveys, and other settings where participants can be given detailed information on the research project up front and can then decide whether or not they wish to accept the disclosed risks and take part.  But while AV testing \textit{riders} can give informed consent before they step into an AV, it is not feasible to obtain similar individual informed consent from every member of the public an AV might encounter (and potentially injure) when it is being tested on public streets.

This difficulty, however, is not unique to AV testing.  Other research areas have confronted similar problems in the past, however, and they have developed systems for obtaining appropriate substituted consent \cite{saver1996critical, carnahan1999promoting}. For example, emergency medical services often provide treatment to individuals in pre-hospital emergency settings who are unconscious, delirious, or otherwise incapable of giving actual consent.  It is thus not possible to obtain informed consent for many EMS research protocols, such as the use of new devices and medical techniques that may pose new risks. The federal Informed Consent Requirements in Emergency Research waiver takes the position that neither forbidding emergency research nor assuming blanket consent is appropriate \cite{oprr1996informed}. Instead, research protocols are evaluated on a case by case basis, where an institutional review board (IRB) approves the research subject to safeguards like ``consultation ... with representatives of the communities in which the research will be conducted and from which the subjects will be drawn'' and ``public disclosure of sufficient information following completion of the research to apprise the community and researchers of the study ... and its results.''  (Similar language is codified in the FDA's informed-consent rules for clinical investigations at 21 C.F.R. §~50.24.)  For AV testing, approval by public authorities satisfies many of the community-wide process values required by the Emergency Research waiver.  Data sharing with those bodies is an important part in their being able to perform this ethically legitimating function. To give properly informed consent (albeit at the community rather than the individual level), they must be properly informed. 

\textbf{Beneficence}, which implements a utilitarian principle of minimizing harm and maximizing benefits, requires researchers to carry out research in ways that minimize risk and to pursue only projects whose benefits from increased knowledge outweigh the harms of the research itself.  Disclosure plays important roles in both.  Because numerous companies are engaged in AV testing, information on the \textit{risks} of such testing is important for minimizing the harms caused by simultaneous and redundant research.  Dangerous road situations are dangerous regardless of which particular AV encounters them: to withhold such information from others conducting AV research makes their research more dangerous than necessary.

There is a useful analogy to the Food and Drug Administration's various adverse-effect reporting requirements.  For example, the FDA requires medical device manufacturers and facilities such as hospitals to submit information on deaths, injuries, or serious malfunctions, both before and after approval. Both the database of reports and individual reports themselves are made publicly available. \cite{adverseeffect}  This regulatory data is thus available both to regulators and to competitors; the process is designed to improve the general level of knowledge about device safety.  Accelerating (so to speak) the development of safer AV technology by companies in general is a way of realizing the principle of beneficence.

Finally, \textbf{justice} implements principles of distributive justice, which require that the benefits and burdens of research be distributed fairly.  Here, AV testing imposes the burden of risk on those members of the public who are in proximity to the AVs being tested, while the primary benefits of increased knowledge are realized by the companies carrying out the testing.  This mismatch between public risk and private benefit is in tension with the principle of justice.

In the scientific domain, publication of research results is a common way of resolving this tension: sharing the knowledge gained from research with the public at large reduces the concern that research participants are being exploited for private gain.  With AV testing, the long term potential benefit to the general public is huge; the widespread deployment of commercial AVs promises to make roads safer.  In the interim, the public bears only risk while the technology is still being developed.  Data sharing is a way of moving  safety benefits forward, so that they more closely compensate the people who bore the risk to make the benefits possible.

\section{What Data is Available to Share?}
AVs and especially AV test vehicles process and record a large variety of data that are used to control the vehicle. 
In this section, we will give an overview of what kind of data is collected and can be synthesized by AVs during test deployments. 

\subsection{Sensory data}
\label{sensorydata}
Sensor data is the low-level information coming directly from sensors attached to the AV. The most common sensors are.

\begin{description}[leftmargin=0cm]
\setlength\itemsep{0.25em}
\item[GPS] Autonomous cars use information from global position satellites to locate their latitude and longitude to about $\leq4.8m$  accuracy \cite{GPSAccurrat}. 

\item [RADAR] Some AV manufacturers use RADAR technology to detect objects and obstacles on the roadway around the vehicle \cite{dungan2010classifying}. RADAR can be used to detect objects multiple car lengths ahead by using their reflections around physical occlusions \cite{Scheiner_2020_CVPR}. 

\item [LIDAR] LIDAR is similar to RADAR in that it is used to capture objects and obstacles on and around the roadway \cite{hamieh2020lidar}. The use of laser-based technology enables greater resolution but can also suffer different shortcomings compared to RADAR technology. 

\item [Cameras] RGB cameras sense information in the visual space using inexpensive CMOS technology, similar to what can be found in computers and cellphones \cite{goberville2020analysis}. Infrared cameras can also ``see'' in the dark by reflecting light in the non-visible spectrum \cite{geng2020using}. Often, multiple cameras are used in stereo to enable modeling of depth \cite{8959696}.

\item [Map Data] The map is the reference against which all other sensor data is matched. Map data is mostly made of pre-existing data and not recorded on the fly. However, although we think of roadways as fixed, they are incredibly dynamic and need to be constantly re-mapped and updated \cite{wang2020offline}. Automated vehicles driving along the road capture and tag geolocated information about temporary or permanent road closures, changes in road conditions, typical traffic on a roadway, or changes in the stores or facilities \cite{chen2019incremental}.

\item[IMU] Inertial Measurement Units only provide relative updates on changes in velocity. However, they do so at a very high rate (>=10KHz), and can therefore be used to interpolate between absolute measures. Many sensors like these are often distributed across the vehicle.
\end{description}

While the sensor streams listed here are the most commonly used, autonomous vehicle are also often used as platforms for novel experimental sensors like millimeter wave distance sensors \cite{kong2017millimeter} or  AI-enhanced microphones \cite{ListeningAV}. 

\subsection{Modeled Data}
\label{modeleddata}
Some information collected by vehicles is not merely sensed, but \textit{inferred}. That is to say, some level of processing or integration of data is used to produce the information. Data from the same sensor might be modelled very differently by different companies, who often develop their own proprietary algorithms and datasets.

\begin{description}[leftmargin=0cm]
\setlength\itemsep{0.25em}
\item [Object detection] Objects, like people, bicycles, other cars, trees, or curbs, need to be ``recognized.'' Raw data from a source like RADAR, LIDAR or cameras need to be interpreted to distinguish that there is an object, and what that object is. Aliasing, where one object can seem like another object, or false-positive detection, where an object is perceived where there is none, occurs with frequency with each sensing technology, so multiple sensing technologies are often  used to increase a system's confidence that it has recognized and correctly identified an object. Ultimately, the systems stores the location, size and relevant characteristics of each object detected. \cite{9000872}

\item [Sign detection] As a special sub-class of object detection, AVs also need to spot and interpret signs along the roadway. These include static signs, like stop signs or road signs, but also dynamic signs, like stop lights, as well as temporary signs, like road construction signs. Interpretation of these signs can be simple when weather conditions are good, and the AV has a head-on view. Bad weather, various occlusions and different viewing angles can make detection of signs more difficult \cite{bos2020autonomy,chincholkar2019traffic}.

\item[Trajectories] For all relevant vehicles and other traffic agents, AVs not only note the location but the trajectory of the vehicle. It is important to distinguish whether vehicles or other road users are moving fast or slow, if they are moving in a direct or erratic function, and whether there are other obstacles or agents that are likely to affect that agent's impending movements. 

\item [Road condition] Data from the autonomous vehicle sensors can be used to map the topology, road markings, and additional features that are placed on the road that help guide traffic. Some characteristics of the roadway may be as designed, but other features, such as potholes or curb deterioration, might be emergent. 

\item[Environmental conditions] Inferences about external conditions, for instance, about weather (raining, hailing), roadway situations (roadside accident, double-parked delivery trucks), human events (parades, protests) can be difficult for AVs to infer now, but represent the complex scene understanding capability that trully autonomous vehicles will need to operate independently.

\end{description}

\subsection{Logged Data}
\label{loggeddata}
Logged data is distinct from sensed data in that it records events or actions of and around the AV. The car might log that it braked, for example. Sometimes the data is automatically logged, other times the safety driver inside the vehicle log data manually.
\begin{description}[leftmargin=0.2cm]
\setlength\itemsep{0.25em}
\item[Engagements/Disengagements] AVs keep track of when the vehicle is being operated by the autonomous system and when they is being operated by the human safety driver, and, specifically, when the human safety driver takes over control of vehicle operation outside of plan. This distinction can be important to determining the cause of faults in the case of accidents. It can also be useful for understanding how often safety drivers have to intervene in the vehicle operation, either because the road conditions or contexts have changed so that the vehicle is no longer in its operational design domain, or because the vehicle is not responding to the existing conditions in a way that the safety driver feels confident about.

\item[Failures] Safety drivers also need to log failures of automation even when there are not immediate consequences to the driving performance; when vehicles lose sight of or mis-estimates its current location, for example, or when the car ``sees'' objects that the human safety driver can see are not there. Failures can occur in sensing, but also in action; sometimes the car fails to go when a light turns green, or drives at the wrong speed for the current context. Safety drivers are usually asked to log these instances even if they do not have to intervene to take control of the system because they can be indications of bugs or more consequential failures in sensing, modelling or planned action.

\item [Ego Vehicle Data] AVs log information about their state, such as heading, speed, fuel rate, engine speed, oil temperature, steering angle, or brake state.

\item[Miles driven] Information about failures or disengagements are not particularly meaningful unless they are mapped to a rate of failure or intervention. Hence, the number of miles driven in automation is important information.
 
\end{description}

\subsection{Aggregate data}
For each type of data, aggregates of that data could be shared. For instance, while images may be captured of every pedestrian on the street encountered by an AV, the shared data could be the count of pedestrians encountered on a given street at the time the vehicle passed. Aggregation like this can help mask trade secrets and personally identifying information, while still providing important information.

\section{How is AV Data currently being shared?}
\label{sec:how}




To explore how data can be collected and shared, we compare how data is shared in two international and two U.S. domestic AV deployments. These were selected from the limited set of cases where public explanation of data sharing has been made available, and were deliberately selected to provide concrete examples of current-day data sharing that illustrate the range of detail, timeliness, and accessibility that occur when governments and companies are left to negotiate data sharing for AVs without policy guidance.


\subsection{Singapore}
In Singapore, test vehicles were first trialed on lightly used roads that were highly instrumented. Once vehicles were permitted, AV testing was limited to a specific geo-fenced area. More importantly, all AVs being tested in the Singapore had to be outfitted with a data recorder capable of storing information in the digital format approved by Singapore's Ministry of Transport. \cite{singapore2017roadtraffic, singapore2022lta} This ``data storage system for automated driving'' (DSSAD) references Organisation Internationale des Constructeurs d'Automobiles' working standard for DSSAD \cite{oica2018dssad}, which specifies that the device records and stores a set of data (``timestamped flags'') during the automated driving sequences of any level 3-5 AV. The data needs to notate if the driver or the system was requested to be in control of the driving task, and who was actually performing the driving task whenever a significant safety-related event occurs. 

The provision of a standard DSSAD can allow the Ministry of Transport to ensure that the data collected by each AV is in a usable format, and can help prevent concerns that the AV company is hiding or obscuring relevant data. In addition, the requirement that any AVs limit their testing to areas which the government has instrumented also increases the chance that the government would be able to independently verify information collected from the DSSADs, to know if information is missing or tampered with.

\subsection{Sweden}
In contrast, in Sweden, the Swedish Transport Agency requires companies applying for a permit to describe their own requirements for control, communication, reporting and evaluation. The agency issues two permits, one for trial operation, and another for the vehicle operation. Companies running trials need to demonstrate that their vehicles meet an acceptable level of safety, but the criteria for the sufficient safety are specified by the applicants themselves. Applicants are obliged to report any incidents, but are given a year to make the report, and of course the report is with the data that the company decide are relevant to share. There is a mandate for a safety driver, but the driver can be in or outside of the vehicle. \cite{swedish2017government,swedish2019permission} This has made it possible to test driverless trucks on public highways \cite{clevenger2019unmanned}, and self-driving Volvos on the streets of Gothenburg \cite{volvo2018swedish}.

\subsection{California}
California requires AV companies to submit annual reports of all the instances where AVs being tested on public roads experience ``collisions'' and ``disengagement.'' The California DMV defines a disengagement as ``a deactivation of the autonomous mode when a failure of the autonomous technology is detected or when the safe operation of the vehicle requires that the autonomous vehicle test driver disengage the autonomous mode and take immediate manual control of the vehicle.'' \cite{CA_DMV2022} The intent of the law is clearly to make it possible for the government and citizens to be able to monitor the safety of the testing programs, but does not clearly define key terms, such as the time frame for the ``immediate'' manual control and ``safe operation.''

\subsection{Massachusetts}
The Commonwealth of Massachusetts permits the testing and deployment of highly automated driving technologies \cite{MA_AV2016} on state highways or other public or publicly accessible state roadways with the approval of the Massachusetts Department of Transportation (MassDOT). MassDOT's application requires information describing the entity’s track record of testing, both on-road and off-road, and including any crash-related information; the results of any relevant safety assessment; information regarding any vehicles to be tested on the public ways; and, significantly ``the sharing of non-proprietary information generated during testing with the AV Working Group.''  Submitted reports have provided some good information 
\cite{bos2020autonomy}), especially in the ``takeovers'' and ``learning'' sections, where they list specific conditions like fog or where oncoming vehicles or bicycles violated lane boundaries, where their AVs were facing challenges.


\subsection{Reflections}

These above examples, though not comprehensive, illustrate the range in substance, detail and format, of the reporting municipalities and governments require of AV companies. 

In the case of Singapore, the government requires that AV companies use data recording equipment that captures information in the exact format that they specify, allowing them to dictate which datastreams are collected, at what rate, and how the data is reported. In contrast, Sweden leaves it up to the AV company to determine what information is relevant to report, and makes hardly any demands for reporting. 

The Swedish requirements are that safety situations need to be reported after the fact, but permits a lot of time---one year---between event and reporting. California's requirement to report both collisions but also disengagements give the government more insight into factors that might turn into accidents with a larger and more widespread deployment. Still, definitions of key terms are left open to interpretation. 

The format of the reports is also important. California's accident reports rely primarily on a narrative description of the facts around the disengagement, and there is no standard for the quality of these descriptions. 
These are publicly available, and the AV company must include information on weather, lighting, roadway surface, roadway conditions, movement preceding collision, type of collision, and any other associated factors.  Many of Waymo’s descriptions read ``unwanted movement of the vehicle that was undesirable under the circumstances''--this is vague to the point of being useless to further analysis. The Massachusetts crash reports are similarly qualitative, narrative, and very unspecific. The descriptions of ``takeovers'' are broad--``in certain situations in which construction vehicles were obstructing our lane of travel.''  
Submitted disengagement and crash reports also do not include key contextual parameters needed to interpret the safety or reliability of the technologies under test--how many miles the companies are covering, how big the fleets are, the number of disengagements that occurred over a period of time, and where the disengagement and take overs occurred. 

The spareness of these disengagement and crash reports is extraordinary mostly because autonomous driving is enabled primarily through the collection of vast amounts of high-resolution data, none of which the state requires be shared. A detailed description of the facts and circumstances that caused the disengagement is the most important factor in making insightful recommendations and group learning; the vagueness of the submitted crash reports makes clear that AV companies will not be forthcoming with any data that is not specifically required of them.

An additional note: the data from both California and Massachusetts is at least highly accessible; reports from both states is available on the state government websites. The degree of availability of the data, however, may motivate some of the obfuscation we note in the filings. The sharing of the ``non-proprietary information generated during testing'' required by Massachusetts, in contrast, is only required to be shared with the state's Autonomous Vehicle Working Group. It is possible that larger but still limited pool of disclosure would make it possible to increase transparency and accountability in the operation of autonomous vehicles while limiting concerns about the disclosure of sensitive proprietary details of a company's AV deployment.

The fact that the reports are submitted in written, paragraph form is another issue, as all the reports submitted by AV companies can look completely different \cite{massgov2016}. Such data is not standard, nor is it machine readable, making it difficult and time consuming for researchers to find the relevant information.



\section{How should AV data be shared?}
Based on the public benefit rationale for data sharing that we established in section \ref{sec:why}, we propose a standard of \textit{proportional information disclosure}, wherein the kind and amount of data that should be shared should lie in proportion to the utility of that data for public benefit.

\subsection{Proportional information disclosure}

We propose that different degrees of data collected by autonomous vehicles should be shared based on situational factors: 

\begin{description}[leftmargin=0.2cm]
\item [Nominal operation] In nominal operation, we expect autonomous vehicles to be good road citizens, and enough data should be provided to understand how the vehicle is interacting with other vehicles and road users. Also, feedback to the municipality about conditions or on-road and near-road events should be reported. The provided data should produce similar value as cameras monitoring a given AV test track. Hence, it should give insight into the safety of the testing operation and provide basic contextual information about the environment. 

\item [Disengagement event] Disengagement events indicate situations where safety drivers either anticipated or noticed the autonomous vehicle would not be able to handle the event. The details surrounding a disengagement event should give a detailed picture about the circumstances such that it can be reconstructed by other testing companies to verify their own software against this failure case. 

\item [Collision] Since collisions must be well understood in order to prevented, the maximal amount of data should be required around any collision event. The main goal of collision reports even for normal drivers is to get at the cause of the incident. AV collision reports should use this ideal and fully exploit the available data stored on the system to enable reconstruction of the incident, context and causal factors. 
\end{description}

With this proposal, there is some amount of data being collected by autonomous cars under normal conditions which are still accorded to the benefit of general public. For example, pedestrian counts are valuable to the commerce department of a city. Uneven street conditions are important for the transportation department in planning infrastructure repair. 

Some data that is being shared relates to safety. It is important not only that AV companies transparently report their own issues with regard to safety, but also share enough information that each additional AV company does not need to recreate unsafe conditions in the public roadway to also learn from and avoid those situations. This admittedly trades away some competitive advantage from companies that perform tests, but for the purpose of minimizing risk to the general public. 




\subsection{Data Standards and Formats}
The data format for sharing these data streams should be open and standardized so that it can be written and read without the use of proprietary software. 

Recent efforts to standardize measurements in autonomous vehicle simulation might well provide a model for real-world autonomous vehicle data sharing. The Association for Standardization of Automation and Measuring Systems (ASAM) is carrying forward the development of openSCENARIO \cite{openscenario}, which is intended to provide a format that specifies simulated events and scenarios in a format that is portable from one simulation platform to another. 

Additionally, ISO  has a working group that is specifically concerned with developing ``Test scenarios of automated driving systems'' \cite{iso_2020} as part of the standard on ``Vehicle dynamics and chassis components.'' 

Data types and formats as developed by the ASAM and ISO working groups should form the basis for the required data-sharing system. The coding of on-road activities of actual autonomous cars using these test scenarios would go a long way toward standardizing reporting of various events and making it easier to recognize the prevalence or absence of various issues in different locales or with different autonomous vehicle deployments. The unification in the specification and coding of simulation and test scenarios to on-road incidents would increase compatibility with existing tool chains, which in turn would lower the barrier to publish disengagements and ease in the testing of scenarios published by other companies.



\subsection{Timely publication of data}
All generated data should be submitted within a fixed window of time after an event. For this time window, we again suggest to follow a proportional approach based on the importance and severity of the information to be reported.

\begin{description}[leftmargin=0.2cm]
\item [Nominal operation]  Data that is regularly generated should also be shared in such a timely, near to real-time manner. This is mostly because findings generated from this information (e.g. traffic congestion around a construction-site) are time critical.
\item [Disengagement events and Collisions] Any particular incident needs to be shared quickly so that all involved can learn from a critical situation and avoid it in the future. And so, depending on the complexity of the incident, data sharing should be available one to two weeks after the accident. This ensures that other testing companies can avoid the same mistake or that the local traffic safety can resolve potentially hazardous traffic environments. 
\end{description}

\subsection{Third-party data sharing management}

The question of \textit{what entity} should perform the data sharing is also important. Currently, data is often shared by the autonomous vehicle company or the regulatory agency where the field deployments occur. This can make it difficult to get a broader picture for what issues are with autonomous vehicle deployments across the country or the world. 

In their 2016 report, the NHTSA recommended following the Federal Aviation Administration's model of using pre-market approval processes to regulate the safety of autopilot software and unmanned systems of aviation vehicles. The Aviation Safety Information Analysis and Sharing (ASIAS) entrusts the not-for-profit MITRE company to collect and share data about aviation safety issues and violations with any requestees.  \cite{business_aviation_insider_2017} This use of a third-party can for logistical reasons--the company can help make sure data is compliant and submitted regularly--but it can also help to make sure that the proportional disclosure of the data does not take place in a self-interested way, and help to prevent uses of the data for unintended purposes.

This use of a third-party mechanism to facilitate anonymous data sharing helps to make sure data is shared in ways that do not violate antitrust and competition, but still restrict access to parties with an appropriate interest in the data. Because some parties should have access to all the data--for example for safety reasons--but all parties should have access to some of the data, some entity needs to make determinations about which parties have which access to the data. 

The EU recently passed legislation which mandated that \textit{all} vehicles be equipped with event data recorders (EDRs) to capture the status of a vehicle and its systems with the purpose of better understanding car accidents. \cite{eu_edr} Böhm et al. suggested mechanisms for third-parties to manage and host the data, and share it with relevant parties. \citeauthor{KubjatkoPaulaSchweiger} They suggest that the third party data management company have the following responsibilities:
\begin{itemize}
    \item reconstructing virtual scenarios of the accident
    \item anonymizing the scenario
    \item sharing the anonymized data through an accident database
\end{itemize}


\subsection{Regulatory oversight}
The aggregation of AV data also makes regulatory oversight and specifications possible. For this reason, all the collected data, be it daily updates or crash data, should be combined into one database of traffic scenarios which should be reviewed regularly by regulatory agencies at all levels of government.

A database of AV driving data should be used as a resource to certify new AV technologies as well as help the general public understand the limitations of this technology. 

\section{Legal issues with AV data sharing}
\label{sec:legal}
Because deployments of autonomous vehicles require a lot of investment, AV companies are likely to challenge data sharing requirements as they have challenged other regulations that require them to disclose information to the government. 
In this section, we consider the legal issues with AV data sharing within the US, which has substantial precedent of preserving corporate interests, and EU, which has substantial legislation regulating the data collected about private persons. 
\subsection{United States}
\subsubsection{Privacy}

AV passengers have legitimate and important privacy interests.  The locations to and from which they travel, for example, can be highly sensitive: for example, to a domestic violence shelter.  In 2019, Uber filed a lawsuit and temporary restraining order against the L.A. Department of Transportation over their mobility data specification (MDS), which required the collection and sharing of trip data from dockless scooters and bikes. The complaints were brought on the basis of user privacy. \cite{bliss2019uber}   In an analogous case, Airbnb successfully argued that a New York City ordinance requiring it to disclose ``voluminous data'' about hosts to the City violated the Fourth Amendment's prohibition on unreasonable searches and seizures.\cite{2019airbnb} 

Fortunately, data sharing requirements on AV safety incidents can be crafted narrowly enough to avoid raising privacy issues.  Because of the focus on failures and near misses, there is no need to disclose trip-level information about starting points and destinations or identifiable information about particular passengers.  This minimizes or eliminates the private information collected.  At the same time, the focus on failures and near misses means that these are situations in which the disclosed data is particularly important to public safety.  Just as the US' National Transportation and Safety Board investigates aviation and surface accidents, and drug and medical device manufacturers are required to report adverse events to the US Food and Drug Administration, data collection on incidents involving danger to human life is reasonable even if it is regarded as a search.

\subsubsection{Trade Secrets}

AV companies have trade secret rights under state and federal law over the data their vehicles generate.  This data has economic value which derives in part from the fact that it is not generally known by competing companies.  Trade secret law generally prohibits the acquisition of a trade secret through ``improper means'' or the knowing use of a trade secret that was obtained through improper means. \cite{utsa}  These rights, however, will not stand in the way of data-sharing requirements.

First, AV companies will not be able to sue under trade secret law itself.  The federal Defend Trade Secrets Act (DTSA) does not apply to ``any otherwise lawful activity conducted by a governmental entity of the United States, a State, or a political subdivision of a State.'' \cite{dtsa}  In \textit{Fast Enterprises v. Pollack}, a federal court held that the DTSA could not be used to prevent the public release under Massachusetts public record law of confidential documents submitted in response to an RFP. \cite{2018fast} Even if they could sue, required disclosures as part of a regulatory scheme are not improper means of acquiring a trade secret; instead, they are a ``lawful means of acquisition.''  \cite{dtsa}

\subsubsection{Takings}

The more difficult argument against data-sharing requirements is that they might constitute a taking of private property.  The Takings Clause of the Fifth Amendment to the United States Constitution states, ``nor shall private property be taken for public use, without just compensation.''  The Takings Clause does not forbid government takings; it just requires that the owner be compensated.  Importantly, it does not just apply to explicit eminent-domain takings, as when the government takes private land to build a highway.  Instead, it also applies to \textit{regulatory takings}, in which ``government regulation of private property may, in some instances, be so onerous that its effect is tantamount to a direct appropriation.''\cite{2005lingle}

The threshold question is whether data is legally considered property, and consequently whether the Takings Clause extends to data at all.  The Supreme Court has answered this question in the affirmative. \textit{Ruckelshaus v. Monsanto Co.} held that because confidential information shares ``many of the characteristics of more traditional forms of property,''  it is subject to the Takings Clause as long as it satisfies the legal definition of a protectable trade secret.  \cite{1984ruckelshaus}

The substantive question, therefore, is whether data-sharing requirements constitute a regulatory taking of AV companies' property rights in their confidential information.  

In our analysis, data-sharing requirements must be tested against the 
three-part ``Penn Central'' test announced in \textit{Penn Central Transp. Co. v. New York City}.\cite{1978penn}.  This test requires consideration of three factors: the regulation’s economic impact on the claimant, the regulation’s interference with their investment-backed expectations, and the character of the governmental action.\cite{echeverria2005making}

Under the first factor, the greater the economic impact of the government action, the more likely it will be considered a ``taking.'' In this context, AV companies have clearly invested large sums of money into perfecting their systems and collecting the data necessary to do so. Having to disclose crash, disengagement, and near-miss scenario data to their competitors could lower the competitive edge they hoped to gain through their investment and possibly lead to economic consequences. However, the data we are asking AV companies to disclose is peripheral -- it contains no details about the content or details of their algorithms. We have also recommended that the data of all companies be aggregated and anonymized, lowering the possibility or magnitude of economic impact even more. Additionally, property under this first prong is viewed as a whole. AV companies are clearly still able to use the disclosed data to improve their own systems even if it is no longer secret, which therefore means it still holds value for the company. This factor seems to favour the government. 

The second factor involves looking at whether the governmental action was reasonably foreseeable when the claimant purchased their property. In \textit{Appolo Fuels, Inc. v. United States}, the Federal Circuit listed three factors that bear on foreseeability: whether the claimant was operating in a ``highly regulated industry'', whether the claimant was aware of the problem the regulation tries to solve when they purchased the property, and whether they could have ``reasonably anticipated'' such a regulation in the context of the ``regulatory environment'' when they made their purchase.\cite{2004appolo} Take these one by one. First, the AV space (and the automobile industry more generally) is heavily, indeed pervasively, regulated. It is illegal to operate a vehicle on public roads unless it meets a lengthy list of engineering standards and unless its operator is property licensed.  Second, AV companies are clearly aware of the safety problems associated with testing their vehicles on public roads, as well as the importance of having robust training data to minimise accidents. Finally, the NHTSA has been requiring crash reports on incidents of all severity since 1988\footnote{https://cdan.nhtsa.gov/tsftables/tsfar.htm}, California implemented a mandatory disengagement disclosure requirement in 2014, and the DOT and NHTSA published their voluntary data sharing guidelines in 2016. 

AV companies have not strictly speaking ``made their purchase'' because there was no purchase -- the data is generated and recorded on an ongoing basis by the AV companies rather than purchased from another party. But however the timing is defined, the safety risks of AVs were readily apparent then, and AV companies were clearly aware that extensive and heavily regulated safety testing would be required.  While AV companies may try to argue that they could not have anticipated data-sharing requirements when they first entered the AV space and started R\&D or manufacturing vehicles or trialling AVs, it is important to remember that the relevant property here is the data generated by AVs during testing.  While backward-looking data-sharing requirements on past incidents might put this final subfactor in play, a forward-looking data-sharing requirement on future incidents does not even implicate it, because the relevant ``purchase'' is the generation of the data in a regulatory environment requiring disclosure.  
Even as to backwards-looking requirements, there is a relevant regulatory history: automotive crash reports.  To be sure, these are both less detailed and occur less frequently than what we are proposing.  But there is also a strong analogy to FAA standards for aircraft, which involve highly  comprehensive reporting requirements in a directly adjacent industry.  This factor also favors the government.

The third factor considers four aspects of the governmental action -- whether it involves a physical occupation of private property, whether it impairs the right to devise private property to heirs, whether it targets specific parties or has a more general application, and whether it is benefit-conferring or harm-preventing.\cite{echeverria2005making} There is obviously no physical occupation here and AV companies remain free to transact freely in the data.  The regulation has a general application to all companies that conduct AV testing and is not intended to restrict any specific company, and the regulation is harm-preventing as it is designed to protect the public as a whole from preventable AV accidents. Based on this analysis, the third factor also favors the government. 

In conclusion, all three of the \textit{Penn Central} factors favor the government.  Data-sharing requirements, as we propose them, would not constitute a taking under the Fifth Amendment, and no compensation would be due to AV companies.  In a 2016 decision, the California Public Utilities Commission applied the \textit{Penn Central} factors to conclude that a trip-level data reporting requirement for an AV operator was not an unconstitutional taking.  \cite{calpuc}

\subsubsection{Unconstitutional Conditions}

One final argument AV companies might raise is that requiring disclosure of testing data as a condition of testing approval would be an \textit{unconstitutional condition}.  In \textit{Phillip Morris v. Reilly}, for example, the court held that a Massachusetts law requiring cigarette manufacturers to disclose ingredient lists was an unconstitutional condition on their right to sell cigarettes.  \cite{2002philip}  The usual test on such conditions is that it is unconstitutional ``to require a person to give up a constitutional right ... in exchange for discretionary benefit conferred by the government where the benefit sought has little or no relationship to the property.''  \cite{1994dolan}

The contrast between \textit{Monsanto} and \textit{Phillip Morris} here is instructive.  \textit{Monsanto} involved a ``complex regulatory scheme'' that otherwise forbade the sale of potentially dangerous pesticides and required disclosure specifically about the components that might make them dangerous.  On the other hand, in \textit{Phillip Morris} Massachusetts otherwise generally allowed the sale of cigarettes, and required disclosure quite broadly, whenever that disclosure ``could reduce risks to public health.''  The AV testing data disclosure requirements we propose here are directly connected to the hazards created by AV testing itself, which is a tightly regulated process that is not otherwise legal.  It involves a clear government benefit in exchange for the disclosure, and the disclosure requirement has a tight nexus to the benefit being granted.  We therefore believe that such requirements are not unconstitutional conditions, either.

\subsection{European Union}
\subsubsection{Privacy}
The General Data Protection Regulation (GDPR)\cite{eu2018gdpr}, drafted and passed in the European Union (EU), went into effect in 2018. The aim of the regulation is to protect individuals in the EU regarding the collecting and processing of their personal data. The origins of privacy legislation in the EU date back to the 1950 European Convention on Human Rights\cite{eu2010convention}. Article 8, 'right to respect for private and family life', states that:
\begin{displayquote}
1. Everyone has the right to respect for his private and family life, his home and his correspondence.

2. There shall be no interference by a public authority with the exercise of this right except such as is in accordance with the law and is necessary in a democratic society in the interests of national security, public safety or the economic well-being of the country, for the prevention of disorder or crime, for the protection of health or morals, or for the protection of the rights and freedoms of others.
\end{displayquote}
With the European Convention on Human Rights as the basis for legislation, technological advances called for updated policies. The European Data Protection Directive \cite{eu1995directive} was passed in 1995, which stated fundamental data privacy and security guidelines, upon which EU member states based their legislation. The GDPR can be seen as the most recent addition of strict data protection regulations in response of increased data collection through technologies and services.

\subsubsection{Fundamental principles of GDPR}
GDPR applies to \textit{personal data}, that is, data relating to an individual (i.e. the \textit{data subject}) who can be directly or indirectly identified from the data. Actions performed on data, such as collecting, structuring, organizing, using, storing and erasing, are captured under the term \textit{data processing}. Next to the data subject, other key parties are the \textit{data controller} (i.e. who decides the reason for and method of data processing) and the potential \textit{data processor} (i.e. a third party who is processing the data on behalf of a data controller).

The data controller is responsible for the fact that data processing happens in according to seven principles (see Art. 5): (1) Lawfulness, fairness and transparency; (2) Purpose limitation; (3) Data minimization; (4) Accuracy; (5) Storage limitation; (6) Integrity and confidentiality; (7) Accountability.
Data controllers are required to prove that they are complying with GDPR, which involves meeting a set of requirements. GDPR further requires data controllers to follow \textit{data protection by design and default} (Art. 25), to ensure that technical and organisational measures are implemented for adhering to GDPR principles and minimization of data collection to the purpose of collection. 

In order to collect data lawfully, at least one of the following conditions must be true (Art. 6):
(1) The data subject provided unambiguous consent; (2) Processing is necessary for the performance of a contract to which the data subject is party in; (3) Processing is necessary for compliance with a legal obligation (that the data controller has); (4) Processing is necessary to protect the life of the data subject or another natural person; (5) Processing is necessary to carry out a task in the public interest; (6) The data processor has a legitimate interest in processing the data subject's data.

The rights of the data subject are described in Chapter 3 (Art. 12-23), which are: (1) right to be informed, (2) right of access, (3) right to rectification, (4) right to erasure, (5) right to restriction of processing, (6) right to data portability (including the right to receive personal data concerning them in a machine readable format, and to distribute this to another party without hindrance from the previous data controller, Art. 20), (7) right to object and (8) rights related to automated decision making and profiling. 

\subsubsection{AV Data and Public Interest}
The previous summary on GDPR principles gives an impression of what is considered fair in the spirit of GDPR. As GDPR regulations are currently limited to data subjects that reside in the EU, they do not apply to all cases of on-road AV testing, especially when testing takes place in other regions of the world. However, we believe that these principles point to a concept of fairness that should be striven for when it comes to the costs and benefits of data processing from the perspective of the \textit{data subject}.

When AVs collect data on public roads, the data will contain information about individuals, who, in most cases, have not explicitly given consent to be test subjects. The balance of costs and benefits is thus already shifted, and there will be additional cost to data subjects in the sense that they are exposed to potentially dangerous situations that relate to AV testing. Due to the nature of AV testing, most data will have been collected in public spaces, where generally speaking, there are no strict regulations on data collection - a tourist, for instance, can take video recordings of a public road, without it being a privacy issue. However, technological advances in data analysis (such as in computer vision), are increasingly compromising the anonymity of individuals, as small details regarding their gait and physique, in combination with the location and time of data collection, can be analyzed to give away a lot of information regarding their identity. AV testing involves systematic collection of large amounts of data, where it cannot be ensured that the data processing will eventually benefit the data subject (for instance, if they have no intention of purchasing a self-driving car).

There needs to be a clear benefit of the data collection to the data subject, which is separate from the purpose of the AV testing entity. In the spirit of GDPR, a data subject has the right to obtain the data collected about them, and provide this to another data processor. As AV data may feature countless individuals which are only indirectly identifiable, it may not be feasible to track down the identities of all data subjects and provide the data to them. However, in a democracy, (local) governments act as representatives and advocates for their residents. We thus argue that (local) governments, should the public wish for it, should have the right to access AV testing data collected in their area of legislation from other data controllers, provided that this data is used to the benefit of the public. 

\section{Discussion}


The contribution of this paper is an argument for the sharing of data collected about the public by privately owned autonomous vehicle companies. This type of data collection goes hand-in-hand with the development and testing of a new technology in public space, and with risk to the public. Because the sharing of this data helps to make the risks and benefits of the technology transparent to the public, and because data sharing reduces redundant testing by competing autonomous vehicle companies, we argue that data should be shared in proportion to the utility of the data to the public, that the data standards and formats be unified so as to make it easier to analyze, that the data be released in a timely fashion, and that a third party---analogous to ASIAS--be used to collect and disseminate the data.

Research on AVs has been prominent in CHI because it raises key issues at the intersection of human factors and technology \cite{kun2016shifting, ayoub2019from}. However, the HCI community should also be advocating for this type of data sharing for research reasons. In this section, we specifically highlight the benefits of data sharing for researchers, benefits of data sharing for the larger community, and future directions for action.

\subsection{Benefit for researchers}

Data from AV deployments would benefit researchers by forecasting impacts of technology, identifying innovation needs, helping to ground research to real-world scenarios and situations, and increasing trust with local communities.

\textbf{Forecasting impact, Increasing innovation} Increasingly, private companies like Meta have started voluntarily sharing data to the academic research community to help social media researchers understand their company's impact on society \footnote{\url{https://fort.fb.com/researcher-platform}}. This approach to data sharing is offered post-facto; as researchers, we only have an opportunity to uncover the consequences of large scale decisions that private companies are effecting on our lives. Advocating for the sharing of data from test deployments, as we are proposing for AVs, allows researchers to analyze and influence the impact that AVs will have on society at a more opportune juncture.

Sharing logged data from AVs (see Section \ref{loggeddata}) about problem areas would help to highlight opportunities for new research. Currently, issues encountered by AV companies are known primarily to those companies; the data reports provided about safety incidents are not detailed enough to infer causation. The disparity in knowledge between researchers in the AV companies and outside make identifying blind spots and finding new and varied avenues of discovery difficult. However, when critical data is shared from AV research (i.e. the details of how an AV car crash occurred), the AV research community is aware of the conditions that created this accident and can take action to avoid inadvertently recreating these conditions.

\textbf{Grounding research to real-world data} 
Modelled data (see Section \ref{modeleddata}) from AV deployments can be beneficial to HCI research because it can help researchers to recreate scenarios and contexts for research based on real-world situations. Currently university researchers often use simulated scenarios to develop their driving algorithms (e.g., CARLA Simulator \cite{Dosovitskiy17}) because more realistic data is expensive or unavailable. The provision of geo-spatial and sensory data (see Section \ref{sensorydata} about objects, road features, or environmental conditions might also be helpful for urban HCI research \cite{fischer2012urban, alavi2019introduction}. 

\textbf{Increased trust with local communities} Mistrust of technologies being foisted on communities has the potential to contaminate other community-based technology research; conversely, the sharing of data with the public has the potential to foster trust between researchers and the communities that they conduct research in.  An understanding of AV technology is correlated with with favorable perceptions of the technology \cite{cmuAVs}. Shared data will help to provide an explanation for the introduction of retrofitted cars in these communities and inform the community what kind of data is being collected.

\subsection{Applications for this data beyond AV testing}


The immediate use case for this data is in oversight of AV testing. Given the current setup with self-reported accident reports and promotional videos, AV developing companies share very little information about how they're developing and implementing safety critical infrastructure. This makes an honest risk analysis close to the impossible. More granular data would give municipalities and research a chance to better understand how this technology will play a role in the wider environment, how rules of the road are interpreted, etc.

A secondary use case for such granular pedestrian and road data is in having high-quality data on road usage. Municipalities currently employ their methods in collecting data on road usage. For example, NYC uses real-time traffic cameras and San Francisco tracks congestion in the city. However, as we discussed in Section 4, AV companies collect data that municipalities do not already collect. The robust datasets can be used for endeavors such as city and road planning and even understand how people move through a city. 

\subsection{Future work for the HCI community}
The HCI community can aid the cause of datasharing from AVs through work in the following areas:

\textbf{Standardizing data formats} A wide variety of data is collected during AV research and this could vary further depending on the standards set by the companies that conduct the research. In order for the data to be consistent across different companies and locations, standard data formats must be agreed upon.

\textbf{Creating mechanisms for sharing} Currently, AV companies that decide to share their data have their own methods of doing so. For example, Waymo hosts their Open Dataset on their own website \cite{waymodata}. The methods of sharing should be standardized to ensure both the availability and accessibility of data.

\textbf{Taking part in policy advocacy} Parts of the CHI community already focus on ethics, law and policy, but researchers focusing on technical issues or human participant studies might not always see the relevance of major policy decisions that are taking place at different levels of their government on their own work. Large scale datasharing has the power to be transformative; the influence of sites like Kaggle \footnote{\url{https://www.kaggle.com/}} and Hugging-face \footnote{\url{https://huggingface.co/}} on the machine learning community show how shared data can advance research with a community. By advocating for datasharing, CHI researchers are advocating for greater safety and transparency, and also for the provision of data which can provide enormous secondary benefit to their own community.





\section{Conclusion}
In this paper, we have made the case that data collected by autonomous vehicles as part of on-road testing and mapping should be shared with the general public, as a form of compensation and control for the risks to communities where such data collection occurs. While we have focused our discussion of the legal and policy implications of such maneuvers on the United States, we believe that the ethical case for such sharing, as well as the pragmatic logistics of how such sharing would work, would be applicable no matter the specific country or region. Because many of the companies that are developing AV technology are multi-national corporations, there may be the tendency for those corporations to do regulatory ``jurisdiction shopping,'' to test and collect data where there are the least demands put upon them to share. Hence, we hope that this work helps members of the CHI community make the argument the idea that data from autonomous vehicles should ethically be shared with the public, so AV data sharing policies will be taken up by governments, municipalities and regulatory bodies which interact with autonomous vehicle technologies on behalf of their people.



 \bibliographystyle{ACM-Reference-Format}
 \balance
 \bibliography{mainBib,acmart,website}


\begin{thebibliography}{85}


\ifx \showCODEN    \undefined \def \showCODEN     #1{\unskip}     \fi
\ifx \showDOI      \undefined \def \showDOI       #1{#1}\fi
\ifx \showISBNx    \undefined \def \showISBNx     #1{\unskip}     \fi
\ifx \showISBNxiii \undefined \def \showISBNxiii  #1{\unskip}     \fi
\ifx \showISSN     \undefined \def \showISSN      #1{\unskip}     \fi
\ifx \showLCCN     \undefined \def \showLCCN      #1{\unskip}     \fi
\ifx \shownote     \undefined \def \shownote      #1{#1}          \fi
\ifx \showarticletitle \undefined \def \showarticletitle #1{#1}   \fi
\ifx \showURL      \undefined \def \showURL       {\relax}        \fi
\providecommand\bibfield[2]{#2}
\providecommand\bibinfo[2]{#2}
\providecommand\natexlab[1]{#1}
\providecommand\showeprint[2][]{arXiv:#2}

\bibitem[way({[n.\,d.]})]%
        {waymodata}
 \bibinfo{year}{[n.\,d.]}\natexlab{}.
\newblock \bibinfo{title}{Waymo Open Dataset}.
\newblock
\newblock
\urldef\tempurl%
\url{https://waymo.com/open/download}
\showURL{%
\tempurl}


\bibitem[197(1978)]%
        {1978penn}
 \bibinfo{year}{1978}\natexlab{}.
\newblock \showarticletitle{{Penn Central Transportation Co. v. New York
  City}}.
\newblock \bibinfo{journal}{\emph{U.S. Supreme Court}} \bibinfo{volume}{438},
  \bibinfo{number}{No. 77-444} (\bibinfo{year}{1978}), \bibinfo{pages}{Page:
  104}.
\newblock


\bibitem[198(1984)]%
        {1984ruckelshaus}
 \bibinfo{year}{1984}\natexlab{}.
\newblock \showarticletitle{Ruckelshaus v. {Monsanto Co.}}
\newblock \bibinfo{journal}{\emph{U.S. Supreme Court}} \bibinfo{volume}{467},
  \bibinfo{number}{No. 83-196} (\bibinfo{year}{1984}), \bibinfo{pages}{Page:
  986}.
\newblock


\bibitem[uts(1985)]%
        {utsa}
 \bibinfo{year}{1985}\natexlab{}.
\newblock \showarticletitle{{Uniform Trade Secrets Act}}.
\newblock \bibinfo{journal}{\emph{Uniform Law Commission}} (\bibinfo{date}{8}
  \bibinfo{year}{1985}).
\newblock


\bibitem[199(1994)]%
        {1994dolan}
 \bibinfo{year}{1994}\natexlab{}.
\newblock \showarticletitle{{Dolan v. City of Tigard}}.
\newblock \bibinfo{journal}{\emph{U.S. Supreme Court}}  \bibinfo{volume}{512}
  (\bibinfo{year}{1994}), \bibinfo{pages}{Page: 374}.
\newblock


\bibitem[200(2002)]%
        {2002philip}
 \bibinfo{year}{2002}\natexlab{}.
\newblock \showarticletitle{{Philip Morris, Inc. v. Reilly}}.
\newblock \bibinfo{journal}{\emph{Court of Appeals, 1st Circuit}}
  \bibinfo{volume}{312}, \bibinfo{number}{F.3d} (\bibinfo{year}{2002}),
  \bibinfo{pages}{Page: 24}.
\newblock


\bibitem[200(2004)]%
        {2004appolo}
 \bibinfo{year}{2004}\natexlab{}.
\newblock \showarticletitle{{Appolo Fuels, Inc. v. US}}.
\newblock \bibinfo{journal}{\emph{Court of Appeals, Federal Circuit}}
  \bibinfo{volume}{381}, \bibinfo{number}{No. 03-5088} (\bibinfo{year}{2004}),
  \bibinfo{pages}{Page: 1338}.
\newblock


\bibitem[200(2005)]%
        {2005lingle}
 \bibinfo{year}{2005}\natexlab{}.
\newblock \showarticletitle{Lingle v. {Chevron USA Inc.}}
\newblock \bibinfo{journal}{\emph{U.S.Supreme Court}} \bibinfo{volume}{544},
  \bibinfo{number}{No. 04-163} (\bibinfo{year}{2005}), \bibinfo{pages}{Page:
  528}.
\newblock


\bibitem[dts(2016)]%
        {dtsa}
 \bibinfo{year}{2016}\natexlab{}.
\newblock \showarticletitle{{Defend Trade Secrets Act}}.
\newblock \bibinfo{journal}{\emph{Stat.}}  \bibinfo{volume}{130}
  (\bibinfo{year}{2016}), \bibinfo{pages}{Page: 376}.
\newblock


\bibitem[201(2018)]%
        {2018fast}
 \bibinfo{year}{2018}\natexlab{}.
\newblock \showarticletitle{{Fast Enterprises, LLC v. Pollack}}.
\newblock \bibinfo{journal}{\emph{Dist. Court, D. Massachusetts}}
  \bibinfo{number}{No. 16-cv-12149-ADB} (\bibinfo{year}{2018}).
\newblock


\bibitem[201(2019)]%
        {2019airbnb}
 \bibinfo{year}{2019}\natexlab{}.
\newblock \showarticletitle{{Airbnb, Inc. v. City of New York}}.
\newblock \bibinfo{journal}{\emph{Dist. Court, S.D. New York}}
  \bibinfo{volume}{373}, \bibinfo{number}{F. Supp. 3d} (\bibinfo{year}{2019}),
  \bibinfo{pages}{Page: 467}.
\newblock


\bibitem[Administration(2016)]%
        {asias_factsheet}
\bibfield{author}{\bibinfo{person}{Federal~Aviation Administration}.}
  \bibinfo{year}{2016}\natexlab{}.
\newblock
\newblock
\urldef\tempurl%
\url{https://www.faa.gov/news/fact_sheets/news_story.cfm?newsId=18195}
\showURL{%
\tempurl}


\bibitem[Administration)(2013)]%
        {nhtsa2013preliminary}
\bibfield{author}{\bibinfo{person}{NHTSA (National Highway Traffic~Safety
  Administration)}.} \bibinfo{year}{2013}\natexlab{}.
\newblock \bibinfo{title}{Preliminary Statement of Policy Concerning Automated
  Vehicles}.
\newblock
\newblock


\bibitem[Administration)(2021)]%
        {nhtsa2021website}
\bibfield{author}{\bibinfo{person}{NHTSA (National Highway Traffic~Safety
  Administration)}.} \bibinfo{year}{2021}\natexlab{}.
\newblock \bibinfo{title}{Automated Driving Systems}.
\newblock
\newblock
\urldef\tempurl%
\url{https://www.nhtsa.gov/vehicle-manufacturers/automated-driving-systems}
\showURL{%
\tempurl}


\bibitem[Agency)(2019)]%
        {swedish2019permission}
\bibfield{author}{\bibinfo{person}{Transport Styrelsen (Swedish~Transport
  Agency)}.} \bibinfo{year}{2019}\natexlab{}.
\newblock \bibinfo{title}{The Swedish Transport Agency's Regulations and
  General Advice (TSFS 2017:92) on Permission to Conduct Trials with
  Self-driving Vehicles}.
\newblock
\newblock
\urldef\tempurl%
\url{https://www.transportstyrelsen.se/en/road/Vehicles/self-driving-vehicles/}
\showURL{%
\tempurl}


\bibitem[Alavi et~al\mbox{.}(2019)]%
        {alavi2019introduction}
\bibfield{author}{\bibinfo{person}{Hamed~S Alavi}, \bibinfo{person}{Elizabeth~F
  Churchill}, \bibinfo{person}{Mikael Wiberg}, \bibinfo{person}{Denis Lalanne},
  \bibinfo{person}{Peter Dalsgaard}, \bibinfo{person}{Ava Fatah~gen Schieck},
  {and} \bibinfo{person}{Yvonne Rogers}.} \bibinfo{year}{2019}\natexlab{}.
\newblock \bibinfo{title}{Introduction to human-building interaction (hbi)
  interfacing hci with architecture and urban design}.
\newblock , \bibinfo{numpages}{10}~pages.
\newblock


\bibitem[ASAM(2021)]%
        {openscenario}
\bibfield{author}{\bibinfo{person}{VIRES Simulationstechnologie~GmbH ASAM}.}
  \bibinfo{year}{2021}\natexlab{}.
\newblock
\newblock
\urldef\tempurl%
\url{http://www.openscenario.org/}
\showURL{%
\tempurl}


\bibitem[Association(2017)]%
        {business_aviation_insider_2017}
\bibfield{author}{\bibinfo{person}{National Business~Aviation Association}.}
  \bibinfo{year}{2017}\natexlab{}.
\newblock \bibinfo{title}{Sharing Aviation Safety Data Is a Good Thing: NBAA -
  National Business Aviation Association}.
\newblock
\newblock
\urldef\tempurl%
\url{https://nbaa.org/aircraft-operations/safety/statistics/sharing-aviation-safety-data-good-thing/}
\showURL{%
\tempurl}


\bibitem[Authority(2022)]%
        {singapore2022lta}
\bibfield{author}{\bibinfo{person}{Land~Transport Authority}.}
  \bibinfo{year}{2022}\natexlab{}.
\newblock \bibinfo{title}{Autonomous Vehicles}.
\newblock
\newblock
\urldef\tempurl%
\url{https://www.lta.gov.sg/content/ltagov/en/industry_innovations/technologies/autonomous_vehicles.html}
\showURL{%
\tempurl}


\bibitem[Ayoub et~al\mbox{.}(2019)]%
        {ayoub2019from}
\bibfield{author}{\bibinfo{person}{Jackie Ayoub}, \bibinfo{person}{Feng Zhou},
  \bibinfo{person}{Shan Bao}, {and} \bibinfo{person}{X.~Jessie Yang}.}
  \bibinfo{year}{2019}\natexlab{}.
\newblock \showarticletitle{From Manual Driving to Automated Driving: A Review
  of 10 Years of AutoUI}. In \bibinfo{booktitle}{\emph{Proceedings of the 11th
  International Conference on Automotive User Interfaces and Interactive
  Vehicular Applications}} (Utrecht, Netherlands)
  \emph{(\bibinfo{series}{AutomotiveUI '19})}. \bibinfo{publisher}{Association
  for Computing Machinery}, \bibinfo{address}{New York, NY, USA},
  \bibinfo{pages}{70–90}.
\newblock
\showISBNx{9781450368841}
\urldef\tempurl%
\url{https://doi.org/10.1145/3342197.3344529}
\showDOI{\tempurl}


\bibitem[Baker(2016)]%
        {MA_AV2016}
\bibfield{author}{\bibinfo{person}{Charlie Baker}.}
  \bibinfo{year}{2016}\natexlab{}.
\newblock \bibinfo{title}{Massachusetts Executive Order No. 572: To Promote the
  Testing and Deployment of Highly Automated Driving Technologies}.
\newblock
\newblock
\urldef\tempurl%
\url{https://www.mass.gov/executive-orders/no-572-to-promote-the-testing-and-deployment-of-highly-automated-driving}
\showURL{%
\tempurl}


\bibitem[{Baraian} and {Nedevschi}(2019)]%
        {8959696}
\bibfield{author}{\bibinfo{person}{A. {Baraian}} {and} \bibinfo{person}{S.
  {Nedevschi}}.} \bibinfo{year}{2019}\natexlab{}.
\newblock \showarticletitle{Improved 3D Perception based on Color Monocular
  Camera for MAV exploiting Image Semantic Segmentation}. In
  \bibinfo{booktitle}{\emph{2019 IEEE 15th International Conference on
  Intelligent Computer Communication and Processing (ICCP)}}.
  \bibinfo{pages}{295--301}.
\newblock


\bibitem[Bliss(2019)]%
        {bliss2019uber}
\bibfield{author}{\bibinfo{person}{Laura Bliss}.}
  \bibinfo{year}{2019}\natexlab{}.
\newblock \bibinfo{title}{Uber's Beef With L.A. Is Bigger Than Data}.
\newblock
\newblock
\urldef\tempurl%
\url{https://www.bloomberg.com/news/articles/2019-10-29/uber-will-take-l-a-to-court-over-data-privacy}
\showURL{%
\tempurl}


\bibitem[Bort(2019)]%
        {bort2019experiment}
\bibfield{author}{\bibinfo{person}{Julie Bort}.}
  \bibinfo{year}{2019}\natexlab{}.
\newblock \showarticletitle{An insider says Uber's self-driving car is more
  like `a science experiment' than a real car that's capable of driving
  itself}.
\newblock \bibinfo{journal}{\emph{Business Insider}} (\bibinfo{date}{19 4}
  \bibinfo{year}{2019}).
\newblock
\urldef\tempurl%
\url{https://www.businessinsider.com/uber-self-driving-car-is-like-science-experiment-insider-says-2019-4}
\showURL{%
\tempurl}


\bibitem[Bos et~al\mbox{.}(2020)]%
        {bos2020autonomy}
\bibfield{author}{\bibinfo{person}{Jeremy~P Bos}, \bibinfo{person}{Derek
  Chopp}, \bibinfo{person}{Akhil Kurup}, {and} \bibinfo{person}{Nathan Spike}.}
  \bibinfo{year}{2020}\natexlab{}.
\newblock \showarticletitle{Autonomy at the end of the Earth: an inclement
  weather autonomous driving data set}. In \bibinfo{booktitle}{\emph{Autonomous
  Systems: Sensors, Processing, and Security for Vehicles and Infrastructure
  2020}}, Vol.~\bibinfo{volume}{11415}. International Society for Optics and
  Photonics, \bibinfo{pages}{1141507}.
\newblock


\bibitem[Braun et~al\mbox{.}(2018)]%
        {braun2018automotive}
\bibfield{author}{\bibinfo{person}{Michael Braun}, \bibinfo{person}{Florian
  Roider}, \bibinfo{person}{Florian Alt}, {and} \bibinfo{person}{Tom Gross}.}
  \bibinfo{year}{2018}\natexlab{}.
\newblock \showarticletitle{Automotive research in the public space: Towards
  deployment-based prototypes for real users}. In
  \bibinfo{booktitle}{\emph{Adjunct proceedings of the 10th international
  conference on automotive user interfaces and interactive vehicular
  applications}}. \bibinfo{pages}{181--185}.
\newblock


\bibitem[Brown(2017)]%
        {brown2017social}
\bibfield{author}{\bibinfo{person}{Barry Brown}.}
  \bibinfo{year}{2017}\natexlab{}.
\newblock \showarticletitle{The social life of autonomous cars}.
\newblock \bibinfo{journal}{\emph{Computer}} \bibinfo{volume}{50},
  \bibinfo{number}{2} (\bibinfo{year}{2017}), \bibinfo{pages}{92--96}.
\newblock


\bibitem[Brown et~al\mbox{.}(2016)]%
        {FiveProvocations}
\bibfield{author}{\bibinfo{person}{Barry Brown}, \bibinfo{person}{Alexandra
  Weilenmann}, \bibinfo{person}{Donald McMillan}, {and} \bibinfo{person}{Airi
  Lampinen}.} \bibinfo{year}{2016}\natexlab{}.
\newblock \showarticletitle{Five Provocations for Ethical HCI Research}. In
  \bibinfo{booktitle}{\emph{Proceedings of the 2016 CHI Conference on Human
  Factors in Computing Systems}} (San Jose, California, USA)
  \emph{(\bibinfo{series}{CHI '16})}. \bibinfo{publisher}{Association for
  Computing Machinery}, \bibinfo{address}{New York, NY, USA},
  \bibinfo{pages}{852–863}.
\newblock
\showISBNx{9781450333627}
\urldef\tempurl%
\url{https://doi.org/10.1145/2858036.2858313}
\showDOI{\tempurl}


\bibitem[Böhm et~al\mbox{.}(2020)]%
        {KubjatkoPaulaSchweiger}
\bibfield{author}{\bibinfo{person}{Klaus Böhm}, \bibinfo{person}{Tibor
  Kubjatko}, \bibinfo{person}{Daniel Paula}, {and} \bibinfo{person}{Hans-Georg
  Schweiger}.} \bibinfo{year}{2020}\natexlab{}.
\newblock \showarticletitle{New developments on EDR (Event Data Recorder) for
  automated vehicles}.
\newblock \bibinfo{journal}{\emph{Open Engineering}} \bibinfo{volume}{10},
  \bibinfo{number}{1} (\bibinfo{year}{2020}), \bibinfo{pages}{140--146}.
\newblock
\urldef\tempurl%
\url{https://doi.org/doi:10.1515/eng-2020-0007}
\showDOI{\tempurl}


\bibitem[Carnahan(1999)]%
        {carnahan1999promoting}
\bibfield{author}{\bibinfo{person}{Sandra~J Carnahan}.}
  \bibinfo{year}{1999}\natexlab{}.
\newblock \showarticletitle{Promoting medical research without sacrificing
  patient autonomy: legal and ethical issues raised by the waiver of informed
  consent for emergency research}.
\newblock \bibinfo{journal}{\emph{Okla. L. Rev.}}  \bibinfo{volume}{52}
  (\bibinfo{year}{1999}), \bibinfo{pages}{Page: 565}.
\newblock


\bibitem[Carroll and Rosson(2007)]%
        {carroll2007participatory}
\bibfield{author}{\bibinfo{person}{John~M Carroll} {and}
  \bibinfo{person}{Mary~Beth Rosson}.} \bibinfo{year}{2007}\natexlab{}.
\newblock \showarticletitle{Participatory design in community informatics}.
\newblock \bibinfo{journal}{\emph{Design studies}} \bibinfo{volume}{28},
  \bibinfo{number}{3} (\bibinfo{year}{2007}), \bibinfo{pages}{243--261}.
\newblock


\bibitem[Chen(2019)]%
        {chen2019incremental}
\bibfield{author}{\bibinfo{person}{Chen Chen}.}
  \bibinfo{year}{2019}\natexlab{}.
\newblock \bibinfo{title}{Incremental updates of pose graphs for generating
  high definition maps for navigating autonomous vehicles}.
\newblock
\newblock
\newblock
\shownote{US Patent 10,267,635}.


\bibitem[Chincholkar and Kumar(2019)]%
        {chincholkar2019traffic}
\bibfield{author}{\bibinfo{person}{YD Chincholkar} {and} \bibinfo{person}{Ayush
  Kumar}.} \bibinfo{year}{2019}\natexlab{}.
\newblock \showarticletitle{TRAFFIC SIGN BOARD DETECTION AND RECOGNITION FOR
  AUTONOMOUS VEHICLES AND DRIVER ASSISTANCE SYSTEMS.}
\newblock \bibinfo{journal}{\emph{ICTACT Journal on Image \& Video Processing}}
  \bibinfo{volume}{9}, \bibinfo{number}{3} (\bibinfo{year}{2019}).
\newblock


\bibitem[Claybrook and Kildare(2018)]%
        {claybrook2018autonomous}
\bibfield{author}{\bibinfo{person}{Joan Claybrook} {and} \bibinfo{person}{Shaun
  Kildare}.} \bibinfo{year}{2018}\natexlab{}.
\newblock \showarticletitle{Autonomous vehicles: No driver… no regulation?}
\newblock \bibinfo{journal}{\emph{Science}} \bibinfo{volume}{361},
  \bibinfo{number}{6397} (\bibinfo{year}{2018}), \bibinfo{pages}{36--37}.
\newblock


\bibitem[Clevenger(2019)]%
        {clevenger2019unmanned}
\bibfield{author}{\bibinfo{person}{Seth Clevenger}.}
  \bibinfo{year}{2019}\natexlab{}.
\newblock \bibinfo{title}{Unmanned Truck Operates on Public Road in Sweden}.
\newblock
\newblock
\urldef\tempurl%
\url{https://www.ttnews.com/articles/unmanned-truck-operates-public-road-sweden}
\showURL{%
\tempurl}


\bibitem[Commission(2016)]%
        {calpuc}
\bibfield{author}{\bibinfo{person}{California Public~Utilities Commission}.}
  \bibinfo{year}{2016}\natexlab{}.
\newblock \bibinfo{title}{Order Instituting Rulemaking on Regulations Relating
  to Passenger Carriers, Ridesharing, and New Online-Enabled Transportation
  Services}.
\newblock
\newblock
\urldef\tempurl%
\url{{https://docs.cpuc.ca.gov/PublishedDocs/Published/G000/M040/K862/40862944.PDF}}
\showURL{%
\tempurl}


\bibitem[Commission(2022)]%
        {eu_edr}
\bibfield{author}{\bibinfo{person}{European Commission}.}
  \bibinfo{year}{2022}\natexlab{}.
\newblock \bibinfo{title}{Vehicle safety – technical requirements \& test
  procedures for EU type-approval of event data recorders (EDRs)}.
\newblock
\newblock
\urldef\tempurl%
\url{https://ec.europa.eu/info/law/better-regulation/have-your-say/initiatives/12989-Vehicle-safety-technical-requirements-test-procedures-for-EU-type-approval-of-event-data-recorders-EDRs-_en}
\showURL{%
\tempurl}


\bibitem[Creaser et~al\mbox{.}(2015)]%
        {creaser2015teen}
\bibfield{author}{\bibinfo{person}{Janet Creaser}, \bibinfo{person}{Nichole
  Morris}, \bibinfo{person}{Christopher Edwards}, \bibinfo{person}{Michael
  Manser}, \bibinfo{person}{Jennifer Cooper}, \bibinfo{person}{Brandy Swanson},
  {and} \bibinfo{person}{Max Donath}.} \bibinfo{year}{2015}\natexlab{}.
\newblock \bibinfo{booktitle}{\emph{Teen driver support system
  (\uppercase{TDSS}) field operational test}}.
\newblock \bibinfo{type}{{T}echnical {R}eport}. \bibinfo{institution}{Center
  for Transportation Studies, University of Minnesota}.
\newblock


\bibitem[des Constructeurs~d'Automobiles(2018)]%
        {oica2018dssad}
\bibfield{author}{\bibinfo{person}{Organisation~Internationale des
  Constructeurs~d'Automobiles}.} \bibinfo{year}{2018}\natexlab{}.
\newblock \bibinfo{title}{Document No. ITS/AD-14-09: DATA STORAGE SYSTEM FOR
  AUTOMATED DRIVING (DSSAD)}.
\newblock
\newblock


\bibitem[Dey and Terken(2017)]%
        {dey2017pedestrian}
\bibfield{author}{\bibinfo{person}{Debargha Dey} {and} \bibinfo{person}{Jacques
  Terken}.} \bibinfo{year}{2017}\natexlab{}.
\newblock \showarticletitle{Pedestrian interaction with vehicles: roles of
  explicit and implicit communication}. In
  \bibinfo{booktitle}{\emph{Proceedings of the 9th International Conference on
  Automotive User Interfaces and Interactive Vehicular Applications}}. ACM,
  \bibinfo{pages}{109--113}.
\newblock


\bibitem[Dingus et~al\mbox{.}(2006)]%
        {dingus2006100}
\bibfield{author}{\bibinfo{person}{Thomas~A Dingus}, \bibinfo{person}{Sheila~G
  Klauer}, \bibinfo{person}{Vicki~L Neale}, \bibinfo{person}{Andy Petersen},
  \bibinfo{person}{Suzanne~E Lee}, \bibinfo{person}{JD Sudweeks},
  \bibinfo{person}{Miguel~A Perez}, \bibinfo{person}{Jonathan Hankey},
  \bibinfo{person}{DJ Ramsey}, {and} \bibinfo{person}{Santosh Gupta}.}
  \bibinfo{year}{2006}\natexlab{}.
\newblock \bibinfo{booktitle}{\emph{The 100-car naturalistic driving study,
  {Phase} {II}-results of the 100-car field experiment}}.
\newblock \bibinfo{type}{{T}echnical {R}eport}. \bibinfo{institution}{United
  States Department of Transportation - National Highway Traffic Safety}.
\newblock


\bibitem[DMV(2022)]%
        {CA_DMV2022}
\bibfield{author}{\bibinfo{person}{California DMV}.}
  \bibinfo{year}{2022}\natexlab{}.
\newblock \bibinfo{title}{California Code of Regulations, Title 13 - Motor
  Vehicles, Division 1 - Department of Motor Vehicles, Article 3.7 - Testing of
  Autonomous Vehicles ( §§ 227.00 - 227.54)}.
\newblock
\newblock
\urldef\tempurl%
\url{https://www.dmv.ca.gov/portal/file/adopted-regulatory-text-pdf/}
\showURL{%
\tempurl}


\bibitem[Domeyer et~al\mbox{.}(2020)]%
        {domeyer2020interdependence}
\bibfield{author}{\bibinfo{person}{Joshua~E Domeyer}, \bibinfo{person}{John~D
  Lee}, \bibinfo{person}{Heishiro Toyoda}, \bibinfo{person}{Bruce Mehler},
  {and} \bibinfo{person}{Bryan Reimer}.} \bibinfo{year}{2020}\natexlab{}.
\newblock \showarticletitle{Interdependence in Vehicle-Pedestrian Encounters
  and Its Implications for Vehicle Automation}.
\newblock \bibinfo{journal}{\emph{IEEE Transactions on Intelligent
  Transportation Systems}} (\bibinfo{year}{2020}).
\newblock


\bibitem[Dosovitskiy et~al\mbox{.}(2017)]%
        {Dosovitskiy17}
\bibfield{author}{\bibinfo{person}{Alexey Dosovitskiy}, \bibinfo{person}{German
  Ros}, \bibinfo{person}{Felipe Codevilla}, \bibinfo{person}{Antonio Lopez},
  {and} \bibinfo{person}{Vladlen Koltun}.} \bibinfo{year}{2017}\natexlab{}.
\newblock \showarticletitle{{CARLA}: {An} Open Urban Driving Simulator}. In
  \bibinfo{booktitle}{\emph{Proceedings of the 1st Annual Conference on Robot
  Learning}}. \bibinfo{pages}{1--16}.
\newblock


\bibitem[Dungan and Potter(2010)]%
        {dungan2010classifying}
\bibfield{author}{\bibinfo{person}{Kerry~E Dungan} {and} \bibinfo{person}{Lee~C
  Potter}.} \bibinfo{year}{2010}\natexlab{}.
\newblock \showarticletitle{Classifying vehicles in wide-angle radar using
  pyramid match hashing}.
\newblock \bibinfo{journal}{\emph{IEEE Journal of Selected Topics in Signal
  Processing}} \bibinfo{volume}{5}, \bibinfo{number}{3} (\bibinfo{year}{2010}),
  \bibinfo{pages}{577--591}.
\newblock


\bibitem[Echeverria(2005)]%
        {echeverria2005making}
\bibfield{author}{\bibinfo{person}{John~D Echeverria}.}
  \bibinfo{year}{2005}\natexlab{}.
\newblock \showarticletitle{Making Sense of Penn Central}.
\newblock \bibinfo{journal}{\emph{UCLA J. Envtl. L. \& Pol'y}}
  \bibinfo{volume}{23} (\bibinfo{year}{2005}), \bibinfo{pages}{171}.
\newblock


\bibitem[Eitner and Colmer(2020)]%
        {ListeningAV}
\bibfield{author}{\bibinfo{person}{Janis Eitner} {and}
  \bibinfo{person}{Christian Colmer}.} \bibinfo{year}{2020}\natexlab{}.
\newblock \bibinfo{title}{Teaching tomorrow’s automobiles to hear}.
\newblock
\newblock
\urldef\tempurl%
\url{https://www.fraunhofer.de/en/press/research-news/2020/february/teaching-tomorrow-s-automobiles-to-hear.html}
\showURL{%
\tempurl}


\bibitem[EUROPE(2018)]%
        {volvo2018swedish}
\bibfield{author}{\bibinfo{person}{CONNECTED AUTOMATED~DRIVING EUROPE}.}
  \bibinfo{year}{2018}\natexlab{}.
\newblock \bibinfo{title}{Swedish Transport Agency allows use of self-driving
  Volvo cars}.
\newblock
\newblock
\urldef\tempurl%
\url{https://web.archive.org/web/20201230040423if_/https://connectedautomateddriving.eu/mediaroom/swedish-transport-agency-allows-use-of-self-driving-volvo-cars/}
\showURL{%
Retrieved November 25 2020 from \tempurl}


\bibitem[{Feng} et~al\mbox{.}(2020)]%
        {9000872}
\bibfield{author}{\bibinfo{person}{D. {Feng}}, \bibinfo{person}{C.
  {Haase-Schütz}}, \bibinfo{person}{L. {Rosenbaum}}, \bibinfo{person}{H.
  {Hertlein}}, \bibinfo{person}{C. {Gläser}}, \bibinfo{person}{F. {Timm}},
  \bibinfo{person}{W. {Wiesbeck}}, {and} \bibinfo{person}{K. {Dietmayer}}.}
  \bibinfo{year}{2020}\natexlab{}.
\newblock \showarticletitle{Deep Multi-Modal Object Detection and Semantic
  Segmentation for Autonomous Driving: Datasets, Methods, and Challenges}.
\newblock \bibinfo{journal}{\emph{IEEE Transactions on Intelligent
  Transportation Systems}} (\bibinfo{year}{2020}), \bibinfo{pages}{1--20}.
\newblock


\bibitem[Fischer and Hornecker(2012)]%
        {fischer2012urban}
\bibfield{author}{\bibinfo{person}{Patrick~Tobias Fischer} {and}
  \bibinfo{person}{Eva Hornecker}.} \bibinfo{year}{2012}\natexlab{}.
\newblock \showarticletitle{Urban HCI: spatial aspects in the design of shared
  encounters for media facades}. In \bibinfo{booktitle}{\emph{Proceedings of
  the SIGCHI Conference on Human Factors in Computing Systems}}.
  \bibinfo{pages}{307--316}.
\newblock


\bibitem[Food and Administration(2020)]%
        {adverseeffect}
\bibfield{author}{\bibinfo{person}{Food} {and} \bibinfo{person}{Drug
  Administration}.} \bibinfo{year}{2020}\natexlab{}.
\newblock \bibinfo{title}{Medical Device Reporting (MDR): How to Report Medical
  Device Problems}.
\newblock
\newblock
\urldef\tempurl%
\url{https://www.fda.gov/medical-devices/medical-device-safety/medical-device-reporting-mdr-how-report-medical-device-problems}
\showURL{%
\tempurl}


\bibitem[for Protection~from Research~Risks(1996)]%
        {oprr1996informed}
\bibfield{author}{\bibinfo{person}{Office for Protection~from Research~Risks}.}
  \bibinfo{year}{1996}\natexlab{}.
\newblock \bibinfo{title}{Informed Consent Requirements in Emergency Research}.
\newblock
\newblock
\urldef\tempurl%
\url{https://www.hhs.gov/ohrp/regulations-and-policy/guidance/emergency-research-informed-consent-requirements/index.html}
\showURL{%
\tempurl}


\bibitem[for the Protection of Human Subjects~of Biomedical and
  Research(1979)]%
        {belmont1979}
\bibfield{author}{\bibinfo{person}{National~Commission for the Protection of
  Human Subjects~of Biomedical} {and} \bibinfo{person}{Behavioral Research}.}
  \bibinfo{year}{1979}\natexlab{}.
\newblock \bibinfo{booktitle}{\emph{The Belmont Report. Ethical Principles and
  Guidelines for the Protection of Human Subjects of Research}}.
\newblock \bibinfo{type}{{T}echnical {R}eport}.
\newblock


\bibitem[Fridman et~al\mbox{.}(2017)]%
        {fridman2017autonomous}
\bibfield{author}{\bibinfo{person}{Lex Fridman}, \bibinfo{person}{Daniel~E
  Brown}, \bibinfo{person}{Michael Glazer}, \bibinfo{person}{William Angell},
  \bibinfo{person}{Spencer Dodd}, \bibinfo{person}{Benedikt Jenik},
  \bibinfo{person}{Jack Terwilliger}, \bibinfo{person}{Julia Kindelsberger},
  \bibinfo{person}{Li Ding}, \bibinfo{person}{Sean Seaman}, {et~al\mbox{.}}}
  \bibinfo{year}{2017}\natexlab{}.
\newblock \showarticletitle{MIT Autonomous Vehicle Technology Study:
  Large-Scale Deep Learning Based Analysis of Driver Behavior and Interaction
  with Automation}.
\newblock \bibinfo{journal}{\emph{arXiv preprint arXiv:1711.06976}}
  (\bibinfo{year}{2017}).
\newblock


\bibitem[Fridman et~al\mbox{.}(2019)]%
        {fridman2019advanced}
\bibfield{author}{\bibinfo{person}{Lex Fridman}, \bibinfo{person}{Daniel~E
  Brown}, \bibinfo{person}{Michael Glazer}, \bibinfo{person}{William Angell},
  \bibinfo{person}{Spencer Dodd}, \bibinfo{person}{Benedikt Jenik},
  \bibinfo{person}{Jack Terwilliger}, \bibinfo{person}{Aleksandr Patsekin},
  \bibinfo{person}{Julia Kindelsberger}, \bibinfo{person}{Li Ding},
  {et~al\mbox{.}}} \bibinfo{year}{2019}\natexlab{}.
\newblock \showarticletitle{MIT advanced vehicle technology study: Large-scale
  naturalistic driving study of driver behavior and interaction with
  automation}.
\newblock \bibinfo{journal}{\emph{IEEE Access}}  \bibinfo{volume}{7}
  (\bibinfo{year}{2019}), \bibinfo{pages}{102021--102038}.
\newblock


\bibitem[Geng and Yin(2020)]%
        {geng2020using}
\bibfield{author}{\bibinfo{person}{Keke Geng} {and} \bibinfo{person}{Guodong
  Yin}.} \bibinfo{year}{2020}\natexlab{}.
\newblock \showarticletitle{Using Deep Learning in Infrared Images to Enable
  Human Gesture Recognition for Autonomous Vehicles}.
\newblock \bibinfo{journal}{\emph{IEEE Access}}  \bibinfo{volume}{8}
  (\bibinfo{year}{2020}), \bibinfo{pages}{88227--88240}.
\newblock


\bibitem[Gessner(2020)]%
        {gessner_2020}
\bibfield{author}{\bibinfo{person}{Daniel Gessner}.}
  \bibinfo{year}{2020}\natexlab{}.
\newblock \bibinfo{title}{Experts say we're decades from fully autonomous cars.
  Here's why.}
\newblock
\newblock
\urldef\tempurl%
\url{https://www.businessinsider.com/self-driving-cars-fully-autonomous-vehicles-future-prediction-timeline-2019-8}
\showURL{%
\tempurl}


\bibitem[Goberville et~al\mbox{.}(2020)]%
        {goberville2020analysis}
\bibfield{author}{\bibinfo{person}{Nick Goberville}, \bibinfo{person}{Mohammad
  El-Yabroudi}, \bibinfo{person}{Mark Omwanas}, \bibinfo{person}{Johan Rojas},
  \bibinfo{person}{Rick Meyer}, \bibinfo{person}{Zachary Asher}, {and}
  \bibinfo{person}{Ikhlas Abdel-Qader}.} \bibinfo{year}{2020}\natexlab{}.
\newblock \bibinfo{booktitle}{\emph{Analysis of LiDAR and Camera Data in
  Real-World Weather Conditions for Autonomous Vehicle Operations}}.
\newblock \bibinfo{type}{{T}echnical {R}eport}. \bibinfo{institution}{SAE
  Technical Paper}.
\newblock


\bibitem[Halford and Harper(2008)]%
        {halford2008asias}
\bibfield{author}{\bibinfo{person}{Carl Halford} {and}
  \bibinfo{person}{Michelle Harper}.} \bibinfo{year}{2008}\natexlab{}.
\newblock \showarticletitle{Asias: Aviation safety information analysis and
  sharing}. In \bibinfo{booktitle}{\emph{2008 IEEE/AIAA 27th Digital Avionics
  Systems Conference}}. IEEE, \bibinfo{pages}{2--C}.
\newblock


\bibitem[Hamieh et~al\mbox{.}(2020)]%
        {hamieh2020lidar}
\bibfield{author}{\bibinfo{person}{Ismail Hamieh}, \bibinfo{person}{Ryan
  Myers}, {and} \bibinfo{person}{Taufiq Rahman}.}
  \bibinfo{year}{2020}\natexlab{}.
\newblock \bibinfo{booktitle}{\emph{LiDAR Based Classification Optimization of
  Localization Policies of Autonomous Vehicles}}.
\newblock \bibinfo{type}{{T}echnical {R}eport}. \bibinfo{institution}{SAE
  Technical Paper}.
\newblock


\bibitem[Hankey et~al\mbox{.}(2016)]%
        {hankey2016description}
\bibfield{author}{\bibinfo{person}{Jonathan~M Hankey},
  \bibinfo{person}{Miguel~A Perez}, {and} \bibinfo{person}{Julie~A
  McClafferty}.} \bibinfo{year}{2016}\natexlab{}.
\newblock \bibinfo{booktitle}{\emph{Description of the {SHRP} 2 naturalistic
  database and the crash, near-crash, and baseline data sets}}.
\newblock \bibinfo{type}{{T}echnical {R}eport}. \bibinfo{institution}{Virginia
  Tech Transportation Institute}.
\newblock


\bibitem[Inners and Kun(2017)]%
        {inners2017beyond}
\bibfield{author}{\bibinfo{person}{Michael Inners} {and}
  \bibinfo{person}{Andrew~L. Kun}.} \bibinfo{year}{2017}\natexlab{}.
\newblock \showarticletitle{Beyond Liability: Legal Issues of Human-Machine
  Interaction for Automated Vehicles}. In \bibinfo{booktitle}{\emph{Proceedings
  of the 9th International Conference on Automotive User Interfaces and
  Interactive Vehicular Applications}} (Oldenburg, Germany)
  \emph{(\bibinfo{series}{AutomotiveUI '17})}. \bibinfo{publisher}{Association
  for Computing Machinery}, \bibinfo{address}{New York, NY, USA},
  \bibinfo{pages}{245–253}.
\newblock
\showISBNx{9781450351508}
\urldef\tempurl%
\url{https://doi.org/10.1145/3122986.3123005}
\showDOI{\tempurl}


\bibitem[Juhlin(1999)]%
        {juhlin1999traffic}
\bibfield{author}{\bibinfo{person}{Oskar Juhlin}.}
  \bibinfo{year}{1999}\natexlab{}.
\newblock \showarticletitle{Traffic behaviour as social
  interaction-implications for the design of artificial drivers}. In
  \bibinfo{booktitle}{\emph{Proceedings of the 6th World Congress on
  Intelligent Transport Systems (ITS)}} (Toronto, Canada).
\newblock


\bibitem[Kong et~al\mbox{.}(2017)]%
        {kong2017millimeter}
\bibfield{author}{\bibinfo{person}{Linghe Kong},
  \bibinfo{person}{Muhammad~Khurram Khan}, \bibinfo{person}{Fan Wu},
  \bibinfo{person}{Guihai Chen}, {and} \bibinfo{person}{Peng Zeng}.}
  \bibinfo{year}{2017}\natexlab{}.
\newblock \showarticletitle{Millimeter-wave wireless communications for
  IoT-cloud supported autonomous vehicles: Overview, design, and challenges}.
\newblock \bibinfo{journal}{\emph{IEEE Communications Magazine}}
  \bibinfo{volume}{55}, \bibinfo{number}{1} (\bibinfo{year}{2017}),
  \bibinfo{pages}{62--68}.
\newblock


\bibitem[Kun et~al\mbox{.}(2016)]%
        {kun2016shifting}
\bibfield{author}{\bibinfo{person}{Andrew~L Kun}, \bibinfo{person}{Susanne
  Boll}, {and} \bibinfo{person}{Albrecht Schmidt}.}
  \bibinfo{year}{2016}\natexlab{}.
\newblock \showarticletitle{Shifting gears: User interfaces in the age of
  autonomous driving}.
\newblock \bibinfo{journal}{\emph{IEEE Pervasive Computing}}
  \bibinfo{volume}{15}, \bibinfo{number}{1} (\bibinfo{year}{2016}),
  \bibinfo{pages}{32--38}.
\newblock


\bibitem[Leonhardt et~al\mbox{.}(2020)]%
        {iso_2020}
\bibfield{author}{\bibinfo{person}{Thorsten Leonhardt}, \bibinfo{person}{Egbert
  Fritzsche}, \bibinfo{person}{Andrew Dryden}, {and} \bibinfo{person}{Fabiola
  Caragol}.} \bibinfo{year}{2020}\natexlab{}.
\newblock \bibinfo{title}{ISO/TC 22/SC 33 - Vehicle dynamics and chassis
  components}.
\newblock
\newblock
\urldef\tempurl%
\url{https://www.iso.org/committee/5383785.html}
\showURL{%
\tempurl}


\bibitem[Mirnig et~al\mbox{.}(2019)]%
        {mirnig2019insurers}
\bibfield{author}{\bibinfo{person}{Alexander~G. Mirnig}, \bibinfo{person}{Rod
  McCall}, \bibinfo{person}{Alexander Meschtscherjakov}, {and}
  \bibinfo{person}{Manfred Tscheligi}.} \bibinfo{year}{2019}\natexlab{}.
\newblock \showarticletitle{The Insurer's Paradox: About Liability, the Need
  for Accident Data, and Legal Hurdles for Automated Driving}. In
  \bibinfo{booktitle}{\emph{Proceedings of the 11th International Conference on
  Automotive User Interfaces and Interactive Vehicular Applications}} (Utrecht,
  Netherlands) \emph{(\bibinfo{series}{AutomotiveUI '19})}.
  \bibinfo{publisher}{Association for Computing Machinery},
  \bibinfo{address}{New York, NY, USA}, \bibinfo{pages}{113–122}.
\newblock
\showISBNx{9781450368841}
\urldef\tempurl%
\url{https://doi.org/10.1145/3342197.3344540}
\showDOI{\tempurl}


\bibitem[National Coordination Office~for Space-Based~Positioning and
  Timing(2020)]%
        {GPSAccurrat}
\bibfield{author}{\bibinfo{person}{Navigation National Coordination Office~for
  Space-Based~Positioning} {and} \bibinfo{person}{Timing}.}
  \bibinfo{year}{2020}\natexlab{}.
\newblock \bibinfo{title}{GPS Accuracy}.
\newblock
\newblock
\urldef\tempurl%
\url{https://www.gps.gov/systems/gps/performance/accuracy}
\showURL{%
\tempurl}


\bibitem[News(2016)]%
        {mercury_2016}
\bibfield{author}{\bibinfo{person}{Mercury News}.}
  \bibinfo{year}{2016}\natexlab{}.
\newblock \bibinfo{title}{Google self-driving car brakes for wheelchair lady
  chasing duck with broom}.
\newblock
\newblock
\urldef\tempurl%
\url{https://www.mercurynews.com/2016/04/05/google-self-driving-car-brakes-for-wheelchair-lady-chasing-duck-with-broom/}
\showURL{%
\tempurl}


\bibitem[Normark(2006a)]%
        {normark2006enacting}
\bibfield{author}{\bibinfo{person}{Daniel Normark}.}
  \bibinfo{year}{2006}\natexlab{a}.
\newblock \emph{\bibinfo{title}{Enacting Mobility. Studies into the Nature of
  Road-related Social Interaction}}.
\newblock \bibinfo{thesistype}{Ph.\,D. Dissertation}.
  \bibinfo{school}{G{\"o}tesborgs Universitet}.
\newblock


\bibitem[Normark(2006b)]%
        {normark2006tending}
\bibfield{author}{\bibinfo{person}{Daniel Normark}.}
  \bibinfo{year}{2006}\natexlab{b}.
\newblock \showarticletitle{Tending to mobility: intensities of staying at the
  petrol station}.
\newblock \bibinfo{journal}{\emph{Environment and Planning A}}
  \bibinfo{volume}{38}, \bibinfo{number}{2} (\bibinfo{year}{2006}),
  \bibinfo{pages}{241--252}.
\newblock


\bibitem[of~Enterprise and Innovation(2017)]%
        {swedish2017government}
\bibfield{author}{\bibinfo{person}{Ministry of Enterprise} {and}
  \bibinfo{person}{Innovation}.} \bibinfo{year}{2017}\natexlab{}.
\newblock \bibinfo{title}{Government paves the way for self-driving vehicles}.
\newblock
\newblock
\urldef\tempurl%
\url{https://www.government.se/articles/2017/05/government-paves-the-way-for-self-driving-vehicles/}
\showURL{%
\tempurl}


\bibitem[of~Health \& Human~Services(1981)]%
        {commonrule}
\bibfield{author}{\bibinfo{person}{U.S.~Department of Health \&
  Human~Services}.} \bibinfo{year}{1981}\natexlab{}.
\newblock \bibinfo{title}{{Federal Policy for the Protection of Human Subjects
  (Common Rule)}}.
\newblock , \bibinfo{numpages}{part 46}~pages.
\newblock


\bibitem[of~Human~Rights and the Council~of Europe(2010)]%
        {eu2010convention}
\bibfield{author}{\bibinfo{person}{European~Court of Human~Rights} {and}
  \bibinfo{person}{the Council~of Europe}.} \bibinfo{year}{2010}\natexlab{}.
\newblock \bibinfo{title}{European Convention on Human Rights}.
\newblock
\newblock
\urldef\tempurl%
\url{https://www.echr.coe.int/Documents/Convention_ENG.pdf}
\showURL{%
Retrieved May 13 2021 from \tempurl}


\bibitem[of~Massachusetts(2020)]%
        {massgov2016}
\bibfield{author}{\bibinfo{person}{Commonwealth of Massachusetts}.}
  \bibinfo{year}{2020}\natexlab{}.
\newblock \bibinfo{title}{Automated Driving Systems in Massachusetts -
  documents and reports}.
\newblock
\newblock
\urldef\tempurl%
\url{https://www.mass.gov/lists/automated-driving-systems-in-massachusetts-documents-and-reports}
\showURL{%
\tempurl}


\bibitem[of~Transport(2017)]%
        {singapore2017roadtraffic}
\bibfield{author}{\bibinfo{person}{Ministry of Transport}.}
  \bibinfo{year}{2017}\natexlab{}.
\newblock \bibinfo{title}{Road Traffic (Autonomous Motor Vehicles) Rules 2017}.
\newblock
\newblock
\urldef\tempurl%
\url{https://sso.agc.gov.sg/SL/RTA1961-S464-2017?ValidDate=20170824&ProvIds=av-#av-}
\showURL{%
\tempurl}


\bibitem[Parliament and the Council of~the European~Union(1995)]%
        {eu1995directive}
\bibfield{author}{\bibinfo{person}{The~European Parliament} {and}
  \bibinfo{person}{the Council of~the European~Union}.}
  \bibinfo{year}{1995}\natexlab{}.
\newblock \bibinfo{title}{European Data Protection Directive}.
\newblock
\newblock
\urldef\tempurl%
\url{https://eur-lex.europa.eu/legal-content/EN/TXT/?uri=celex\%3A31995L0046}
\showURL{%
Retrieved May 13 2021 from \tempurl}


\bibitem[Parliament and the Council of~the European~Union(2016)]%
        {eu2018gdpr}
\bibfield{author}{\bibinfo{person}{The~European Parliament} {and}
  \bibinfo{person}{the Council of~the European~Union}.}
  \bibinfo{year}{2016}\natexlab{}.
\newblock \bibinfo{title}{General Data Protection Regulation}.
\newblock
\newblock
\urldef\tempurl%
\url{https://eur-lex.europa.eu/eli/reg/2016/679/oj}
\showURL{%
Retrieved May 13 2021 from \tempurl}


\bibitem[Reig et~al\mbox{.}(2018)]%
        {cmuAVs}
\bibfield{author}{\bibinfo{person}{Samantha Reig}, \bibinfo{person}{Selena
  Norman}, \bibinfo{person}{Cecilia~G. Morales}, \bibinfo{person}{Samadrita
  Das}, \bibinfo{person}{Aaron Steinfeld}, {and} \bibinfo{person}{Jodi
  Forlizzi}.} \bibinfo{year}{2018}\natexlab{}.
\newblock \showarticletitle{A Field Study of Pedestrians and Autonomous
  Vehicles}. In \bibinfo{booktitle}{\emph{Proceedings of the 10th International
  Conference on Automotive User Interfaces and Interactive Vehicular
  Applications}} (Toronto, ON, Canada) \emph{(\bibinfo{series}{AutomotiveUI
  '18})}. \bibinfo{publisher}{Association for Computing Machinery},
  \bibinfo{address}{New York, NY, USA}, \bibinfo{pages}{198–209}.
\newblock
\showISBNx{9781450359467}
\urldef\tempurl%
\url{https://doi.org/10.1145/3239060.3239064}
\showDOI{\tempurl}


\bibitem[Riener and Reder(2014)]%
        {riener2014collective}
\bibfield{author}{\bibinfo{person}{Andreas Riener} {and}
  \bibinfo{person}{Johann Reder}.} \bibinfo{year}{2014}\natexlab{}.
\newblock \showarticletitle{Collective data sharing to improve on driving
  efficiency and safety}. In \bibinfo{booktitle}{\emph{Adjunct Proceedings of
  the 6th International Conference on Automotive User Interfaces and
  Interactive Vehicular Applications}}. \bibinfo{pages}{1--6}.
\newblock


\bibitem[Saver(1996)]%
        {saver1996critical}
\bibfield{author}{\bibinfo{person}{Richard~S Saver}.}
  \bibinfo{year}{1996}\natexlab{}.
\newblock \showarticletitle{Critical care research and informed consent}.
\newblock \bibinfo{journal}{\emph{N.C. L. Rev.}}  \bibinfo{volume}{75}
  (\bibinfo{year}{1996}), \bibinfo{pages}{Page: 205}.
\newblock


\bibitem[Scheiner et~al\mbox{.}(2020)]%
        {Scheiner_2020_CVPR}
\bibfield{author}{\bibinfo{person}{Nicolas Scheiner}, \bibinfo{person}{Florian
  Kraus}, \bibinfo{person}{Fangyin Wei}, \bibinfo{person}{Buu Phan},
  \bibinfo{person}{Fahim Mannan}, \bibinfo{person}{Nils Appenrodt},
  \bibinfo{person}{Werner Ritter}, \bibinfo{person}{Jurgen Dickmann},
  \bibinfo{person}{Klaus Dietmayer}, \bibinfo{person}{Bernhard Sick}, {and}
  \bibinfo{person}{Felix Heide}.} \bibinfo{year}{2020}\natexlab{}.
\newblock \showarticletitle{Seeing Around Street Corners: Non-Line-of-Sight
  Detection and Tracking In-the-Wild Using Doppler Radar}. In
  \bibinfo{booktitle}{\emph{The IEEE/CVF Conference on Computer Vision and
  Pattern Recognition (CVPR)}}.
\newblock


\bibitem[{\v{S}}ucha(2014)]%
        {vsucha2014road}
\bibfield{author}{\bibinfo{person}{Mat{\'u}{\v{s}} {\v{S}}ucha}.}
  \bibinfo{year}{2014}\natexlab{}.
\newblock \showarticletitle{Road users’ strategies and communication:
  driver-pedestrian interaction}.
\newblock \bibinfo{journal}{\emph{Transport Research Arena (TRA)}}
  (\bibinfo{year}{2014}).
\newblock


\bibitem[Vinkhuyzen and Cefkin(2016)]%
        {vinkhuyzen2016developing}
\bibfield{author}{\bibinfo{person}{Erik Vinkhuyzen} {and}
  \bibinfo{person}{Melissa Cefkin}.} \bibinfo{year}{2016}\natexlab{}.
\newblock \showarticletitle{Developing socially acceptable autonomous
  vehicles}. In \bibinfo{booktitle}{\emph{Ethnographic Praxis in Industry
  Conference Proceedings}}, Vol.~\bibinfo{volume}{2016}. Wiley Online Library,
  \bibinfo{pages}{522--534}.
\newblock


\bibitem[Wang et~al\mbox{.}(2020)]%
        {wang2020offline}
\bibfield{author}{\bibinfo{person}{Zhen Wang}, \bibinfo{person}{Xiangmo Zhao},
  {and} \bibinfo{person}{Zhigang Xu}.} \bibinfo{year}{2020}\natexlab{}.
\newblock \showarticletitle{Offline mapping for autonomous vehicles with
  low-cost sensors}.
\newblock \bibinfo{journal}{\emph{Computers \& Electrical Engineering}}
  \bibinfo{volume}{82} (\bibinfo{year}{2020}), \bibinfo{pages}{106552}.
\newblock


\end{thebibliography}

\end{document}